\documentclass[pre,preprint,showpacs,preprintnumbers,amsmath,amssymb]{revtex4-2}
\usepackage{graphicx}
\usepackage{dcolumn}
\usepackage{bm}
\usepackage[mathscr]{eucal}
\usepackage{mathrsfs}

\begin{document}

\title{Information, order, complexity, and entropy in materials including biological systems: a thermodynamic theory based on state variables}

\author{Koun Shirai$^{1,2}$}

\affiliation{%
$^{1}$ Department of Precision Engineering, Graduate School of Engineering, The
University of Osaka, 2-1, Yamadaoka, Suita, Osaka 565-0871, Japan.
}%
\affiliation{%
$^{2}$Vietnam Japan University, VNU, Hanoi \\
Luu Huu Phuoc Road, My Dinh 1 Ward, Nam Tu Liem District, Hanoi, Vietnam
}%

\begin{abstract}
Entropy plays a central role in thermodynamics, statistical mechanics, information theory, and biology. However, its interpretation becomes increasingly ambiguous when information-theoretic concepts are applied to materials, including biological systems. For example, in biology, it is common practice to evaluate the entropy of DNA by enumerating possible configurations. This configuration entropy does not vanish at $T=0$, apparently contradicting the third law. Similar conceptual difficulties also arise in relating entropy to order, randomness, complexity, and information. 
In thermodynamics, entropy is a state function, and hence the entropy must be uniquely determined by a given state of a material. The crucial issue is therefore to identify the state variables that uniquely specify the thermodynamic state of a material.
By establishing consistent definition of equilibrium and state variable, it is found that the time-averaged atom positions serve as the state variables of a solid. This leads to the important conclusion that a solid possesses many equilibrium states even at fixed temperature and volume. Entropy is not information but uncertainty associated with the state variables. The latter quantities convey the information of a material.
This framework provides a unified thermodynamic basis of entropy, information, order, complexity, hysteresis, and residual entropy while preserving the third law. Frozen configurations and their activation resolve many longstanding ambiguities the thermodynamic evaluation of entropy.
\end{abstract}

\maketitle

\section{Introduction}
\label{sec:Introduction}
\subsection{Impacts of information theory}
\label{sec:Dilemmas}

Thermodynamics is a difficult subject even for specialists \cite{note-Sommerfeld}. 
Interpretation of entropy is the central difficulty in thermodynamics.

\paragraph{From disorder interpretation to information theory}
Today, a measure of ``disorder" is the standard interpretation of entropy \cite{Callen}. In thermodynamics, the definition of entropy is firmly established. The entropy $S$ of a given system is given by
\begin{equation}
\frac{\partial S}{\partial U} = \frac{1}{T},
\label{eq:ST-T}
\end{equation}
where $U$ is the internal energy of the system and $T$ is the absolute temperature.
Despite the clarity of this thermodynamic definition, the concept of entropy has become controversial in many fields. Problems occur when entropy is connected to information theory. Historically, this connection originated from studies of Maxwell's demon, particularly through the work of Szilard \cite{Szilard29}. Since then, information has come to be regarded as a physical quantity, leading to new concepts such as negentropy, the erasure principle, and algorithmic entropy. Many of seminal papers on this subject are collected in a reprint volume \cite{Leff-Rex2}. Consequently ``missing information" has become a new interpretation of entropy \cite{Ben-Naim}. By defining entropy as a measure of uncertainty associated with information, information theory has been developed \cite{Shannon48,Jaynes57,Jaynes57a,Cover-Thomas,Khinchin,SFworkshop89,Parrondo15}. The information-theoretic entropy $H$, also known as Shannon's entropy, is defined as
\begin{equation}
H = -K \sum_{j} p_{j} \ln p_{j},
\label{eq:HInform}
\end{equation}
where $p_{j}$ is the probability of occurrence of event $j$. Here, $K$ is an appropriate constant in favor of a specific application.

Certainly, the concept of missing information has enriched our understanding of entropy in thermodynamics. Some formulations inspired by information theory have been integrated in the framework of statistical mechanics \cite{Wehrl78,Balian03,Martinelli17,NonextEntropy04}. At the same time, however, it has also given rise to a number of conceptual difficulties. In information theory, probabilities are updated as new observations become available. Whether such an interpretation is appropriate in thermodynamics remains uncertain \cite{Collins06}. Many relevant controversies are collected in a special volume \cite{Capek}. In thermodynamics, the entropy of a material must exist regardless of whether it is observed. The updating interpretation of probability has also led to the long-standing question of whether information is fundamentally subjective or objective. Although this issue has been debated extensively, no generally accepted resolution has yet emerged \cite{Jauch-MD2,Beauragard-MD2,Denbigh81,Rosenkrantz83,Loewer01}.
In addition, a material contains information at many different levels, ranging from the macroscopic to microscopic. No textbook explains which kinds of information should be included in the entropy definition. In information theory, there is generally no need to distinguish different kinds of information \cite{Bavaud09}. There, thermodynamics entropy is regarded as a special case of a more general measure of information uncertainty \cite{Caticha21,Tribus71,Cover-Thomas}.
Thermodynamics, however, imposes much stricter restriction on the entropy. As shown by Eq.~(\ref{eq:ST-T}), thermodynamic entropy is fundamentally related to energy. Information theory has no corresponding energy restriction. In particular, there is no counterpart of temperature in information theory, whereas lack of temperature makes thermodynamics theory no sense. More importantly, {\em the thermodynamic entropy is a state variable}. The significance of this property will become clearer in the subsequent sections. Here, we are concerned exclusively with the thermodynamic entropy $S$ of materials. Accordingly, the single term {\em entropy} refers to thermodynamic entropy $S$ unless otherwise specified. However, the treated materials are not limited to inorganic materials but covers any material in general including biological materials.

\paragraph{Order and information}
In solid state physics, the concept of order is often regarded as self-evident. Order is taken to imply perfect periodicity or geometrical regularity. Any imperfection means a disordered state. The entropy is sometimes decomposed into the vibrational ($S_{\rm v}$) and configurational ($S_{\rm c}$) contributions,
\begin{equation}
S = S_{\rm v} + S_{\rm c}.
\label{eq:S=vib+conf}
\end{equation}
The part $S_{\rm c}$ contains the information how many configurations of defects are included, 
\begin{equation}
S_{\rm c} = k_{\rm B} \ln W_{\rm c},
\label{eq:Sconfig}
\end{equation}
where $W_{\rm c}$ is the number of these configurations and $k_{\rm B}$ is Boltzmann's constant. Within this interpretation, $S_{\rm c}$ quantifies the configurational uncertainty associated with defects. Consequently, materials with highly disordered structures, such as glasses, are expected to possess large entropies even at $T=0$, which are called {\it residual entropies}.

This interpretation of order, however, becomes increasingly problematic as a wider variety of materials is considered. Advances in materials science have revealed solids with structures that cannot be described by conventional crystallographic periodicity, such as incommensurate crystals and quasicrystals. Although these materials do not possess ordinary space-group symmetries, they are recognized as exhibiting some forms of order. It is not easy to determine their entropies by experiment. The extreme case may be biological systems. Their thermodynamic entropies are not straightforward to interpret experimentally.
Biology may be the field that has been most influenced by information theory.
DNA and protein are polymers whose monomer sequences exhibit little geometrical regularity. Nevertheless, few would argue that living systems are highly disordered. Motivated by this observation, Schr{\"o}dinger described living matter as an {\em aperiodic crystal}, remarking that ``every atom is playing there" \cite{Schrodinger44} (p.~82). His visionary idea ultimately inspired the discovery of the double-helical structure of DNA. Since then, his fundamental question ``What is life?" has motivated generations of researchers \cite{Murphy-Oneill97,Kaneko09,Nurse21}. However, the thermodynamic meaning of the seemingly paradoxical expression {\em aperiodic crystal} has remained unclear.

\paragraph{Complexity interpretation}
By leaving the dilemma between order and regularity unresolved, researchers came to recognize another aspect of entropy, namely, {\it complexity}. The primary function of DNA is to transmit biological information accurately from one generation to the next. DNA contains all detailed information required to preserve a given species. From the perspective of information theory, a natural interest is: {\it How much information is stored in the base sequence of a given DNA molecule?} \cite{Gatlin72} (p.~20). A single DNA molecule can encode more than $2 \times 10^{4}$ proteins. 
This question strongly motivates the use of the configuration entropy $S_{\rm c}$ to quantify the information content $I$ \cite{Gatlin72,Brooks86,Wicken87}.
There are subtle differences in interpretation and emphasis of entropy among biological researchers; a comparative discussion of these different viewpoints is given in \cite{Collins06}. However, the common manner of interpretation of entropy in biology may be as follows.

Consider a polymer consisting of $N_{L}$ monomers. There are $N_{\alpha}$ types of monomers. Then, the number of possible configurations is $W_{\rm c} = (N_{\alpha})^{N_{L}}$. This represents the maximum number of configurations available to the polymer and is an extremely large number. The corresponding maximum configuration entropy $S_{\rm cm}$ is regarded as representing the information-storage capacity of the polymer. By assuming that the probability of occurrence of $j$-th type of monomer is the same at every site of the polymer, {\it i.e.}, $p_{j} = p \equiv 1/N_{\alpha}$, we obtain
\begin{equation}
S_{\rm cm} = -k_{\rm B} N_{\rm L} \ln p  
  = -k_{\rm B} N_{\rm L} \sum_{j} p_{j} \ln p_{j}.
\label{eq:Scm}
\end{equation}
Hence, this $S_{\rm cm}$ is equivalent to the information-theoretic entropy, Eq.~(\ref{eq:HInform}), apart from an appropriate choice of normalization constant. 
\footnote{The assumption that the probabilities $p_{j}$ are independent at different sites is equivalent to treating the polymer sequence as a Markov process. Experimental observations indicate that this independence is generally violated. In such cases, correlation entropies among different sites are introduced. This makes the analysis increasingly complicated as the number of correlated sites increases.}
Simple crystals are considered to have no capacity for storing information because of the unique structure. Hence, it is natural to define {\rm complexity} by this maximum configuration entropy. The quantity $S_{\rm cm}$ corresponds to the configuration entropy when the occurrence of monomers is completely random.
In reality, the probabilities of occurrence of the individual monomers, $p_{j}$, are not equal. Therefore, the real configuration entropy $S_{\rm cr}$ evaluated by Eq.~(\ref{eq:HInform}) is always smaller than $S_{\rm cm}$. The difference $S_{\rm cm}-S_{\rm cr}$ is then interpreted as representing the amount of useful information contained in the system. Accordingly, the {\it information content} $I$, is defined as
\begin{equation}
I = S_{\rm cm} - S_{\rm cr} \geq 0.
\label{eq:I=Smr}
\end{equation}
When the distribution is completely random, the system conveys no information. As the real configuration entropy decreases from its maximum value, the information content carried by the system correspondingly increases.

The definition of information-theoretic entropy as the entropy of DNA has already been adopted in vast body of the biological literature. This usage has become so widespread that it is unlikely to be reversed. Despite its widespread acceptance, the author argues in this paper that the use of ensemble average is inappropriate for evaluating the thermodynamic entropy of DNA. An individual DNA molecule possesses only one particular sequence of nucleotides, and this sequence does not vary with time. Therefore, an individual DNA molecule has no configuration entropy. 
\footnote{A few researchers have expressed the view that the entropy of DNA is zero in informal media: for example, J. Gielis, https://medium.com/@johan.gielis/the-entropy-of-dna-is-zero-a486c0f4ebee. It is indeed rare to find this view in regular journals or textbooks.}
The configuration entropy of DNA arises only when the category of DNA is considered as a species  rather than an individual molecule. Thus, the central issue is the attribution of entropy: is entropy a property of an individual or of a species? This attribution problem is investigated deeply in Sec.~\ref{sec:Origin}. There, it is shown that the attribution problem is not specific to biology but is common to all materials. 

Furthermore, the above interpretation of complexity involves a serious inconsistency.
It was argued that biological systems possess great complexity and therefore carry rich information, whereas crystals have simple structures and hence cannot carry information. However, this is not the case.
Although a silicon crystal has the simple diamond structure, myriad kinds of electronic devices are fabricated from a single Si wafer. A 1-GB memory chip can store a large amount of information. The conventional interpretation provides no explanation for this apparent inconsistency.
This inconsistency urges us to reconsider the question: ``What is complexity?" In information theory, there is no unique definition of complexity, as many different concepts have been proposed, including Kolmogorov complexity, algorithmic complexity, statistical complexity even in sociology \cite{Cover-Thomas,Chaitin75,SFworkshop89,Bennett89,Gell-Mann95,Gell-Mann96,Beltrami20,Nicolis-Prigogine89}. 

If we again use the degree of order to characterize the capability of information processing, this interpretation appears to work, because biological systems and silicon share the common feature of being highly ordered. However, this leads to another inconsistency. Experimentalists have long recognized the so-called memory effects in glasses \cite{Vincent07,Kovacs79}, even though glasses appear to possess no order. Memory devices utilizing amorphous state are already commercially available, for example, in DVD technology \cite{Wuttig07}. Multivalued memory instead of binary-valued memory has also been investigated using the amorphous state of phase-change materials \cite{Lankhorst05,Raoux08,Zhang19}.

\subsection{Return to thermodynamics}
\label{sec:ReturnTD}
We have seen above that, starting from the question ``What is life?", a chain of questions arises, ``What is order?", ``What is complexity?", and ``What is information?" Adding a question magnifies the inconsistencies, including those between order and regularity, missing information and order, randomness and order, and complexity and information. Although each interpretation may be suitable within its own field, it often becomes inadequate when applied to other fields. We arrive at a deep impasse. Given the universal applicability of thermodynamic theory, such inconsistencies are clearly undesirable.

The author suspects that research has followed a wrong direction. The motivation of the present study originated from research on the physics of glass. The nature of glass is one of the most difficult problem in solid state physics. Glass is often characterized as a ``frozen liquid" and is generally regarded as a nonequilibrium state. Although extensive studies have been devoted to this subject over a century, general consensus has not been obtained \cite{Kauzmann48,Angell95,Wolynes12,Stillinger13,Berthier16}.
Glasses exhibit no apparent regularity in their atomic structures. Despite this, calorimetric measurements suggest the existence of some kind of ``order". At the glass transition, the specific heat exhibits a jump, $\Delta C_{g}$, which has traditionally been attributed to the presence of additional degrees of freedoms beyond $T$ and volume ($V$). These additional degrees of freedom were termed {\it order parameter}, but their physical nature has long remained unknown \cite{Davies53a,Nemilov-VitreousState}. 
Recent studies of the author's group demonstrated that the origin of the jump, $\Delta C_{g}$, is the change in the internal energy $U$ associated with the structural change, similar to the case of phase transitions in crystalline solids \cite{Shirai22-SH, Shirai23-Silica}. Because glass has traditionally been regarded as a frozen liquid---liquids have no fixed structure, our conclusion was not immediately accepted. However, amorphous solids are essentially {\it aperiodic crystals}. Although they possess no periodic cells, the entire glass can be regarded as a single unit cell of ultralarge dimensions. Unlike liquids, all atoms in a glass occupy well-defined equilibrium positions, $\{ \bar{\bf R}_{j} \}$. Hence, the internal energy of a glass is determined by these equilibrium positions as
\begin{equation}
U = U(T, \{ \bar{\bf R}_{j} \} ).
\label{eq:U-FREsolid}
\end{equation}

In parallel with this study, the author has reached the more general conclusion that the time-averaged atom positions, $\{ \bar{\bf R}_{j} \}$, serve as the state variables for solids \cite{Shirai20-GlassState,Shirai18-StateVariable}. From a microscopic viewpoint, the functional relationship of Eq.~(\ref{eq:U-FREsolid})---which states that the energy is determined by the detailed structure---may appear self-evident. However, when Eq.~(\ref{eq:U-FREsolid}) is interpreted as a thermodynamic relationship, $\{ \bar{\bf R}_{j} \}$ naturally emarge as state variables. This relationship is referred to as {\em the fundamental relation of equilibrium} (FRE) \cite{Callen}. 
This conclusion has led to a number of fruitful developments in the thermodynamic of solids. An examination of the meanings of state variables and order parameters has shown that these two concepts are equivalent \cite{Shirai25-OrderParams}. This equivalence provides the key to resolving the inconsistencies described above. By using Eq.~(\ref{eq:ST-T}), the energy-representation of FRE, Eq.~(\ref{eq:U-FREsolid}), can be rewritten in the entropy representation (\cite{Callen}, p.~40)
\begin{equation}
S = S(T, \{ \bar{\bf R}_{j} \} ).
\label{eq:S-FREsolid}
\end{equation}
When $\bar{\bf R}_{j}$ are interpreted as the order parameters, it becomes clear that entropy and order are distinct quantities. This point is discussed in detail in Sec.~\ref{sec:Info}.
The conclusion that the equilibrium atom positions serve as the state variables of solids has far-reaching consequences for several long-standing problems in the thermodynamic of solids, including hysteresis problem \cite{Shirai26-hysteresis}, the nature of glass transition \cite{Shirai-SH-Liquids25}, and the third-law problem \cite{Shirai22-res}. In this paper, the utility of the FRE of solids, Eq.~(\ref{eq:S-FREsolid}), is further extended to biological systems.
This conclusion about the state variables, however, differs substantially from the conventional view that thermodynamics cannot deal microscopic quantities such as atom positions. 

Thermodynamics was established more than a century ago, primarily through studies of gas systems. The four laws of thermodynamics, from the zeroth to third laws, are, of course, equally valid for solids. However, in the study of solids, the primary interest of physicists gradually shifted toward microscopic theories, particularly statistical mechanics and quantum mechanics. Consequently, thermodynamics has rarely been used as the principal framework for investigating the properties of solids. This historical development led to the widespread belief that thermodynamics is applicable only to macroscopic phenomena and therefore should not involve microscopic quantities. In the author's view, this belief has prevented the identification of state variables for solids. It is correct that the thermodynamic properties of one-component gases are uniquely determined by $T$ and $V$ only. The internal energy of a one-component gas is expressed as
\begin{equation}
U = U(T,V).
\label{eq:FRE-gas}
\end{equation}
This relationship may be referred to as {\it the two-variables assumption}. 
However, the validity of the two-variables assumption has never been established for solids, although it is often taken for granted. Nevertheless, because it is difficult to identify additional ``macroscopic" variables other than $T$ and $V$ for solids, the two-variables assumption has conventionally been extended to solids.
However, nothing in the laws of thermodynamics prohibits the use of microscopic quantities as state variables. 
The first law expresses the energy conservation, which is valid irrespective of the size of the system. The second law establishes the upper bound on the efficiency of heat engines, and this principle is equally valid even for microscopic systems. Today, cooling of microscopic cluster of alkali atoms has attracted considerable interest in connection with the Bose-Einstein condensation. The second law serves as the guiding principle in achieving such unprecedentedly low temperatures. Even the word ``quantum thermodynamics" exists \cite{Vinjanamphathy16,Goold16,Deffiner19}. Thermodynamics also serves as a guiding principle for an extreme matter of black holes, where many familiar laws of physics do not hold \cite{Jacobson96}.

The conventional belief described above has become so deeply rooted that many researchers regard thermodynamics as a mature subject lying outside the frontier of current research. This impression is, however, misleading. Thermodynamics theory has continued to develop since its establishment more than a century ago, although the progress has been gradual and has attracted relatively little attention. Developments particularly relevant to the present study include the following: hysteresis can be described thermodynamically by introducing additional state variables \cite{Bridgman50}; the consequences of thermodynamics and statistical mechanics are equivalent \cite{Tisza63,Mandelbrot64}; there is a one-to-one correspondence between constraints and state variables \cite{Reiss}; the distinction between extensive and intensive quantities is not fundamental \cite{Grandy}; and equilibrium can be defined without reference to state variables \cite{Gyftopoulos}. On the basis of these developments, the above conclusion that the state variables of a solid are the equilibrium positions of all atoms comprising the solid, Eq.~(\ref{eq:U-FREsolid}), has been deduced \cite{Shirai20-GlassState,Shirai18-StateVariable}.
Despite these developments, even authoritative textbook by Callen \cite{Callen} does not well reflect them. Consequently, it is natural that many readers still retain the conventional view. For this reason, the essential arguments leading to Eq.~(\ref{eq:U-FREsolid}) are briefly reviewed again in Sec.~{\ref{sec:TDofSolids}}

\subsection{Purpose and construction of this paper}
The purpose of this paper is to establish a rigorous definition of entropy for solid materials, including biological systems, in order to resolve the fundamental inconsistencies discussed above. It should be emphasized again that the entropy studied here is the thermodynamic entropy $S$. Information-theoretic entropy $H$ lies outside the scope of the present study. However, resolving these inconsistencies inevitably requires us to ask ``What is information?" Information theory provides no unique answer to this question \cite{Bavaud09}. In the author's understanding, this is because information theory is not a theory to study information itself. Rather, information is treated as predefined quantity, while the principal concern is how information can be processed, transmitted, or encoded efficiently. This recognition is shared with the view of the biophysicist Eigen \cite{Eigen71}. In physics, it is difficult to study a concept without first defining it. Therefore, the thermodynamic meanings of information, order, and complexity, in materials should be examined with the same rigor as that applied to entropy.

The paper is organized as follows. 
First, the origin of the problem is analyzed in Sec.~\ref{sec:Origin}. It is shown that the difficulties in describing biological entropy are common problems in other solids and are, in fact, closely related to several long-standing problems in thermodynamics. Through an examination of these problems, it is shown that the central question is the definition of equilibrium for solids, and this section sets the orientation of the following arguments.
In Sec.~\ref{sec:TDofSolids}, starting from an unambiguous definition of equilibrium, we identify the state variables of solids. This completes the functional relationship between entropy and the state variables. 
In Sec.~\ref{sec:FC}, the notion of frozen state variable is introduced. This notion explains why different values can be assigned to the same state of a material. It is shown that meaningful comparisons of entropy between different states require the use of a common thermodynamic space. 
Sections \ref{sec:TDofSolids} and \ref{sec:FC} are the summary of the author's previous work \cite{Shirai22-res}, but are included here because they allow readers to follow the logical development of the theory within the present paper.
In Sec.~\ref{sec:Info}, based on the rigorous definition of state variables, the thermodynamic meanings of order, complexity, and information of material are established. It is shown that these quantities are conceptually distinct from entropy.
In Sec.~\ref{sec:ConfS}, the frozen-variable nature of configuration entropy is explained. Frozen state variables can become active thermodynamic variables under appropriate conditions. This transition explains why the current value of entropy of a material may become history dependent.
Finally, Sec.~\ref{sec:Conclusion} summarizes the present study.

\section{Analysis of the problem: multivalued character of entropy}
\label{sec:Origin}

\subsection{Thermodynamic definition of entropy}
\label{sec:thermodynamic-entropy}
The notion of entropy in thermodynamics is generally regarded as well established, and hence it may seem unnecessary to discuss its definition. However, entropy can be defined in several equivalent ways. The most familiar definition is the Clausius definition: the entropy change from state $1$ to $2$ of a system is given by integrating heat $Q$ absorbed by the system along a path connecting these two states. 
\begin{equation}
\Delta S_{21} = \int_{1}^{2} \left( \frac{dQ}{T} \right)_{\rm rev},
\label{eq:defS}
\end{equation}
provided that the path is reversible. The existence of reversible path connecting these terminal states is assumed. Throughout this paper, the experimental determination of entropy based on Eq.~(\ref{eq:defS}) is referred to as the {\it calorimetric} method.
In the present study, however, the axiomatic approach based on Carath\'{e}odory theory is adopted because our primary concern is the fundamental question, ``What is entropy?" The axiomatic formulation provides the most rigorous framework of thermodynamics and is therefore well suited to the present purpose. Within this framework, entropy is defined as a measure of {\em the adiabatic inaccessibility} \cite{Landsberg56,Buchdahl,Pippard,Landsberg70,Giles70,Lieb99}. 
In more intuitive terms, entropy measures the degree of irreversibility in an adiabatic process. 
The following discussion briefly reviews this concept, following Buchdahl's textbook \cite{Buchdahl}.

A thermodynamic state $\mathfrak{S}$ of a system $A$ is specified by a set of state variables, $\{T, X_{1}, \cdots, X_{M} \}$, as
\begin{equation}
\mathfrak{S} = \mathfrak{S}(T, \{ X_{j} \}),
\label{eq:State-TX}
\end{equation}
where $M$ is the number of independent state variables other than $T$. 
Hereafter, the term ``state" always refers to a thermodynamic state, namely, an equilibrium state.
Carath\'{e}odory theory states that, in the vicinity of any given state $\mathfrak{S}$, there always exists another state $\mathfrak{S}'$ that is adiabatically inaccessible from $\mathfrak{S}$. Accordingly, all states of a given system can be classified according to their adiabatic accessibility. The degree of adiabatic inaccessibility is represented by a single scalar function $\sigma$, called {\it empirical entropy}. At this stage, $\sigma$ is merely an ordering parameter for equilibrium states and has not yet been identified with the thermodynamic entropy.
In order to be able to construct such a scalar function from the family of adiabatic curves, 
\begin{equation}
dU + \sum_{j} F_{j} dX_{j} = 0,
\label{eq:ADcurves}
\end{equation}
an integral factor $\lambda$ must be introduced so that these differential equations become integrable,
\begin{equation}
d \sigma = \frac{dQ}{\lambda}, 
\label{eq:integral-factor}
\end{equation}
where $\lambda$ can be taken as a universal function that is independent of the particular system. 

There is an arbitrariness in the choice of the entropy function $\sigma$. Any function $\sigma^{\ast}$ obtained by a monotonic transformation, $\sigma^{\ast} = g(\sigma)$, where $g$ is an arbitrary monotonic function, is equally acceptable as an entropy function. This arbitrariness is removed by specifying the integrating factor $\lambda$. In this way, $\lambda$ provides the scale of the entropy values and is a function of temperature. The resulting entropy is called the {\em metric entropy} $S$. In modern thermodynamics, the quantity appearing in Eq.~(\ref{eq:ST-T}) is defined to be the thermodynamic absolute temperature. Consequently, the entropy $S$ of a material is uniquely determined by the calorimetric method.
An important consequence of this construction is that the scalar function $S$ is obtained by integrating adiabatic differential equations with respect to all state variables $\{ X_{j} \}$,
\begin{equation}
S = S(T, \{ X_{j} \}).
\label{eq:Sfunction}
\end{equation}
This means that {\it there exists one and only one entropy value for each state}. In other words, {\it entropy is a state function}, just like the internal energy $U$, as noted in Introduction. 
Equation (\ref{eq:Sfunction}) is the entropy representation of FRE, which has already appeared as Eq.~(\ref{eq:S-FREsolid}). Thus, Eq.~(\ref{eq:S-FREsolid}) provides further support for identifying $\{ \bar{\bf R}_{j} \}$ as the state variables of solids.

Equation (\ref{eq:Sfunction}) also indicates that the entropy function depends on the choice of state variables. When one set of state variables is replaced by another set of the same dimension, $M'=M$, the transformation must satisfy a Legendre transformation (\cite{Callen}, p.~137), and the entropy of a given state remains unchanged. In contrast, when a new independent variable is introduced, the dimension of the thermodynamic state space changes, and the entropy assigned to a given physical state may also change. This situation arises in connection with the anthropomorphic nature of entropy (see Sec.~\ref{sec:Anthropomorphism}). The essential problem, therefore, is to identify the appropriate set of state variables $\{ X_{j} \}$ for a given state $\mathfrak{S}$. For gas systems, the two-variables assumption is valid, provided that chemical reactions are absent. For solids, however, this problem is by no means trivial. Unfortunately, Buchdahl, like other authors, does not discuss which variables in addition to $T$ and $V$ should be included in the thermodynamic description of solids.

\subsection{Statistical mechanics interpretation}
\label{sec:SMdefinitionS}
The development of physics has often proceeded by explaining macroscopic quantities in terms of microscopic quantities. In statistical mechanics, entropy of a system composed of $N$ particles is defined by the number of microscopic states, $\Omega$, in the phase space, $\Gamma = \{ {\bf r}^{N}, {\bf v}^{N} \}$. Since microscopic quantities are generally regarded as more fundamental, the statistical-mechanical definition of entropy is often considered to provide a deeper understanding than its thermodynamic counterpart. However, this is not necessarily true. Statistical mechanics admits many entropy definitions, such as relative entropy \cite{Wehrl78,Tsallis23}. Surprisingly, many of them are not equivalent. Among these, two have been studied most extensively and are particularly relevant to the present study \cite{Jaynes65,Penrose79, Lebowitz93,Lebowitz93a,Callender04,Benguigui13,Broeck15,Swendsen17}.
The one is Boltzmann entropy, defined by
\begin{equation}
S_{\rm B}(\mathfrak{S}) = k_{\rm B} \ln \Omega(\mathfrak{S}),
\label{eq:Boltzmann-S}
\end{equation}
for a macroscopic state $\mathfrak{S}$. 
The other widely used definition is Gibbs entropy, 
\begin{equation}
S_{\rm G} = -k_{\rm B} \sum_{i \in {\mathfrak M}} p_{i} \ln p_{i},
\label{eq:Gibbs-S}
\end{equation}
where $p_{i}$ is the probability of microscopic state $i$. In Eq.~(\ref{eq:Gibbs-S}), summation extends over all states in the ensemble space ${\mathfrak M}$, that is, ${\mathfrak M} = \{ {\mathfrak S} \}$. Gibbs entropy is often preferred in information theory because the maximum-entropy principle is naturally formulated in terms of this definition \cite{Jaynes57,Jaynes57a,Cover-Thomas}.
When the ensemble ${\mathfrak M}$ contains only a single macroscopic state $\mathfrak{S}$, we have $p_{i} = 1/\Omega(\mathfrak{S})$, and thus the two definitions give the same answer. This is the case for ideal gases. The difficulty arises in the general case, where the two definitions yield different answers. The natural question is which definition is correct. This issue has been debated for decades and yet remains unresolved. Although Lebowitz expressed a preference for the Boltzmann definition, he ultimately regarded the question as open in his later study \cite{Lebowitz93a}. In practice, physicists adopt either definition according to the physical problem under consideration.

Boltzmann entropy is assigned to an individual macroscopic state and is therefore well suited to describing the time evolution of a simple system. Here, a {\it simple system} refers to the one for which the property of {\em typicality}, in the sense used by Lebowitz, holds. In such a system, the statistical properties of a single particle obtained by time averaging are identical to those obtained from the ensemble average over all particles in the system \cite{Lebowitz93,Lebowitz93a}.
Gibbs entropy, on the other hand, emphasizes the probabilistic aspect of ensemble. In this formulation, the central question is the choice of the ensemble space $\mathfrak{M}$: namely, which microstates should be included in the ensemble average?
This issue becomes particularly acute when the entropy of glass is investigated. When the entropy of a glass sample is measured by calorimetric method, the experiment probes only the present structure of that particular sample. Since the structure does not change during the measurement, the configuration entropy vanishes, i.e., $S_{c}=0$. This corresponds to the viewpoint of Boltzmann entropy. However, the generally-held view in glass research is that a glass possesses a large configuration entropy, so that $W_{c}>0$. This interpretation implicitly assumes an ensemble of possible glass structures and therefore reflects the Gibbs viewpoint. This view raises a conceptual difficulty: how can the entropy of an individual sample depend on the structural configurations that are not realized in that sample? This question can be reformulated as the attribution problem of entropy.
Should the entropy of a glass be attributed to an individual glass sample or to the class (or {\it species}) of glasses to which it belongs? This question, sample or species, is the same kind of question that was met when DNA was discussed, i.e., individual or species. 

In fact, the entropy-attribution problem is very fundamental. It is a problem neither specific to glasses nor to DNA, but is the common to all solids. Consider the entropy of a silicon crystal. The vibration part of entropy, $S_{\rm v}(T,V)$, is uniquely determined by $T$ and $V$, and this part can be ignored in the following discussion. Silicon has a definite crystal structure of diamond type, so that its configuration entropy is usually treated to vanish. However, every Si wafer is not the same. Although silicon wafers used in industries are among the most perfect crystals ever produced, every wafer contains intrinsic defects, such as vacancies. The concentration $c$ of vacancies is typically of the order of $10^{14}$ cm$^{-3}$ \cite{Fahey89,Tan85}. Accordingly, each wafer possesses a configuration entropy, $S_{\rm c} = -k_{\rm B} c \ln c$ per atom, when $c$ is expressed as a fractional concentration. Consequently, different wafers have a different entropy values. Moreover, the vacancy concentration is controlled by manufacturers. Therefore, taking an ensemble average over different wafers has no well-defined physical meaning.
In addition to point defects, there are many kinds of extended defects, such as dislocations, grain boundaries, making the notion of an ensemble average even less meaningful.

At this point, the two-variable assumption is often invoked to avoid this difficulty. This assumption claims that even a solid has only one equilibrium state for a given $T$ and $V$, which must be a defect-free crystal. Defect states are accordingly regarded as nonequilibrium states and are excluded from thermodynamic consideration.
However, this conclusion is not appropriate. At finite temperatures, the equilibrium state is determined not by the minimum internal energy but by the minimum free energy. Owing to the logarithmic dependence of the configuration entropy on the defect concentration $n_{d}$, the free energy of a defect-containing crystal always has a minimum at a finite value of $n_{d}$, corresponding to the equilibrium defect concentration, $n_{d}^{\rm eq}$ (\cite{Kittel-Therm}, Chap.~11). As the name implies, this is an equilibrium state, In this sense, a defect-free crystal is a nonequilibrium states at any temperature higher than 0. Since different types of defects have different equilibrium concentrations, there are many independent equilibrium defect concentrations that characterize a real crystal. It is therefore very difficult to specify a unique equilibrium state for a given crystal. Further details are given in \cite{Shirai26-hysteresis}.

As discussed in Sec.~\ref{sec:thermodynamic-entropy}, thermodynamic entropy is a state variable. Therefore, it is inappropriate to take ensemble average if the ensemble space $\mathfrak{M}$ consists of different thermodynamic states, that is, $\mathfrak{M} = \{\mathfrak{S} \}$.
Notwithstanding, Gibbs entropy should not be dismissed altogether. 
The Gibbs formula becomes indispensable when mixing phenomena are treated \cite{Penrose79}. There is no doubt that the mixing entropy of multicomponent gases and the configuration entropy of random alloys are genuine thermodynamic entropies. For gas mixtures, the mixed state itself is the equilibrium state $\mathfrak{S}$, and therefore it is no problem to use $S_{\rm G}$ because  $S_{\rm G}=S_{\rm B}$. Random alloys require a more careful analysis of equilibrium, in particular whether the mixed state constitutes a single equilibrium state or a combination of distinct equilibrium states. This issue is discussed in Sec.~\ref{sec:ConfS}.

\subsection{Hysteresis}
\label{sec:Hysteresis}
One of the long-standing problems in thermodynamics is the treatment of hysteresis. Hysteresis appears in a wide range of phenomena in solids, including magnetic hysteresis, plastic deformation, memory effects. Nevertheless, it is rarely discussed in thermodynamics textbooks. Hysteresis means that the present properties of a material depend on the past history which it has evolved. For this reason, hysteresis is commonly regarded as a nonequilibrium phenomenon, to which conventional thermodynamics analysis is considered inapplicable. However, this reasoning of nonequilibrium causes a serious consequence for thermodynamics. Plastic deformation is a property shared by virtually all solids. Hence, if hysteresis is regarded as sufficient evidence that a system is in a nonequilibrium state, one would be forced to conclude that all solids are nonequilibrium systems. This conclusion would severely limit the applicability of thermodynamics to solids. Yet this fundamental contradiction has received remarkably little attention.

\begin{figure}[htbp]
    \centering
     \includegraphics[width=80 mm, bb=0 0 450 280]{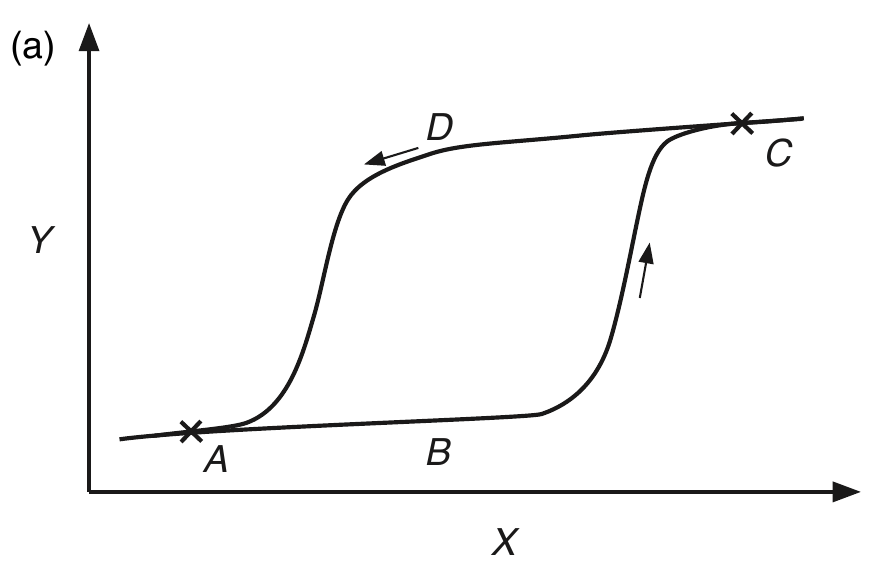} 
     \includegraphics[width=60 mm, bb=0 0 340 280]{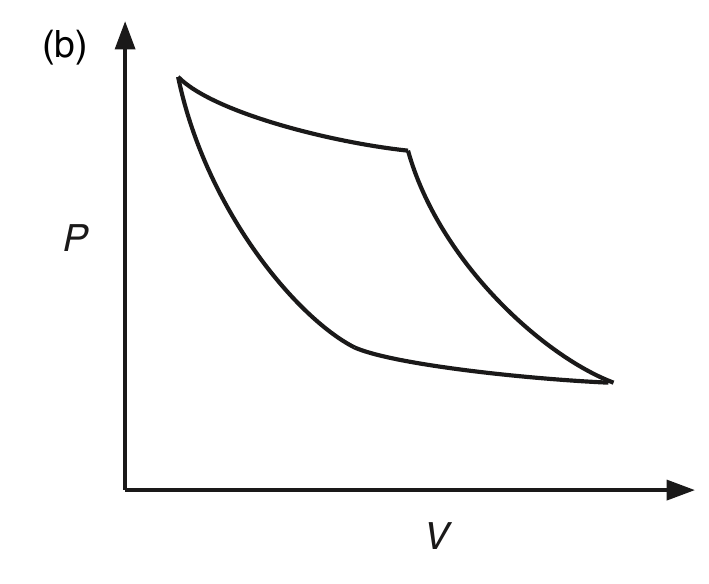} 
  \caption{(a) hysteresis loop. The input variable $X$ and its response $Y$ of a system return to the initial values upon a cycle with different paths for the forward and backward directions of $X$. (b) Carnot cycle plotted on the $V$-$P$ plane. } \label{fig:Hysteresis}
\end{figure}

Figure {\ref{fig:Hysteresis}}(a) shows a schematic hysteresis loop. For a cyclic change of the input variable $X$, the response $Y$ of a material traces a loop. From this, hysteresis is commonly described as a multivalued function \cite{Brokate96}. However, defining hysteresis solely in terms of multivaluedness is incomplete.
Consider the Carnot cycle of an ideal gas. The pressure ($P$) versus volume ($V$) curve also forms a loop, as shown in Fig.~ {\ref{fig:Hysteresis}}(b). Yet, Carnot cycle is the most reversible process, meaning that every state of the gas throughout the cycle is an equilibrium state. This example demonstrates that the multivaluedness does not necessarily imply nonequilibrium.
The apparent multivaluedness arises from the way of representing the thermodynamic process. For an ideal gas, the pressure is a function of two variables, $P=P(V, T)$. During the Carnot cycle, another variable $T$ is also changed. Projecting the two-variable function $P=P(V, T)$ onto a one-dimensional space $V$ yields an apparently multivalued function. This explanation is familiar from standard textbook \cite{Zemansky}. For solids, however, the situation is much less clear, because the appropriate state variables have not been identified. In particular, the widespread acceptance of the two-variables assumption has obscured the resolution of this problem.

Bridgman is one of the few researchers who addressed hysteresis from a thermodynamic perspective \cite{Bridgman50,Bridgman61}. He cautioned against hastily concluding ``the laws of thermodynamics do not apply to a system simply because we are initially unsuccessful in finding a set of state variables which determine the energy function" (Ref.~\cite{Bridgman61}, p.~59). Instead, he suggests that ``we have not a complete list of state variables."
Recently, the author has resolved the hysteresis problem by identifying a complete set of state variables for solids \cite{Shirai26-hysteresis}. The required state variables are the equilibrium positions of all atoms comprising the solid, as expressed by Eq.~(\ref{eq:U-FREsolid}).
From this viewpoint, hysteresis should not be defined in terms of a multivalued function but rather in terms of irreversibility.

\noindent
{\bf Definition of hysteresis}
{\em Hysteresis is a process in which the initial state $\mathfrak{S}_{i}$ of a system is not recovered after one complete cycle of the external field $X$, without causing any change in the environment.}

In Fig.~\ref{fig:Hysteresis}, the terminal states, $A$ and $C$, appear to be uniquely specified by the relation, $Y=Y(X)$. Thus, after the cycle $A \rightarrow B \rightarrow C \rightarrow D \rightarrow A$, one might conclude that the system has returned to its initial state. However, restoration of the initial value for the two displayed variables, namely, $X_{f}=X_{i}$ and $Y_{f}=Y_{i}$, does not guarantee that the final state is identical to the initial state. There may exist another variable $Z$ for which  $Z_{f} \neq Z_{i}$. This is precisely what occurs in hysteresis phenomena. To restore $Z$ to the initial value, additional work must be performed. This necessarily produces a change in the environment, implying that the process is irreversible. This is the essential feature that distinguishes hysteresis from the Carnot cycle. In fact, in hysteresis, although two variables $X$ and $Y$ return to their initial values, the internal structure of the material has changed. Although the crystallographic structure may remain unchanged, microscopic rearrangements, such as microscopic slip, dislocations generation, grain boundary migration, may occur. The formation of even a single defect alters the internal energy $U$, as indicated by Eq.~(\ref{eq:U-FREsolid}).
For further details, refer to the original paper \cite{Shirai26-hysteresis}. 

An important consequence of the study in Ref.~\cite{Shirai26-hysteresis} is that, for a solid, there exist many (almost infinitely many) equilibrium states for a given $T$ and $V$. In a hysteresis loop shown in Fig.~\ref{fig:Hysteresis}, not only the terminal states, $A$ and $C$, but every point on the loop represents an equilibrium states, provided that the process is carried out quasistatically. From modern microscopic theory, once the complete set of equilibrium atom positions, $\{ \bar{\bf R}_{j} \}$, is specified, all static properties of a material are uniquely determined and can, in principle, be reproduced both experimentally and computationally. Reader is encouraged to keep this fact in mind throughout the following discussion.

\subsection{The third law issue}
\label{sec:ThirdLaw}
The attribution problem of entropy---whether entropy is a property of individual sample or of a species---becomes particularly evident in connection with the third law of thermodynamics, which is another long-standing problem in thermodynamics.
Although all window glasses belong to the same glass of materials, having essentially the same chemical compositions and optical properties, each sample possesses a different atom arrangement. Each sample has its own unique structure. From this viewpoint, the configuration entropy of an individual glass sample should vanish at $T= 0$. This is the conclusion reached when Boltzmann entropy is adopted. In contrast, it is generally accepted that a glass has a finite entropy even at $T=0$. This is residual entropy. Theoretically, this residual entropy is interpreted as the configurational entropy $S_{\rm c}$, which is obtained when Gibbs entropy is adopted. The existence of residual entropy appears to violate the third law of thermodynamics. 
The entropy of DNA that is obtained by adopting Gibbs entropy likewise remains finite at $T=0$.
However, this apparent conflict with the third law is rarely discussed in the biological literature. For glasses, the conventional explanation has been that residual entropy arises because glasses are nonequilibrium states. Recently, however, this interpretation has been critically reexamined by the present author \cite{Shirai22-res}. 

\begin{figure}[htbp]
    \centering
    \includegraphics[width=140 mm, bb=0 0 860 340]{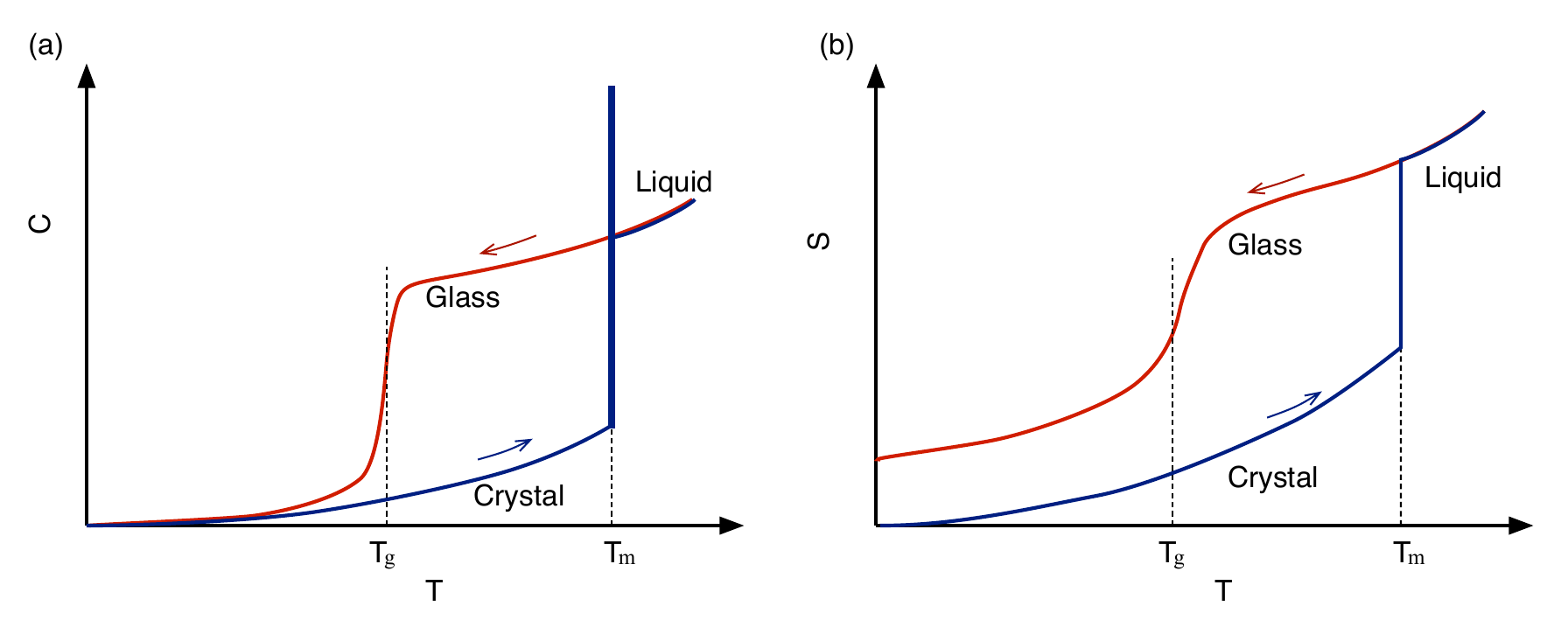} 
  \caption{Specific heat ($C$) and entropy ($S$) of a glass measured during a cooling process in calorimetry.
  } \label{fig:CSofGlass}
\end{figure}

Aside from the question of whether the Boltzmann or Gibbs definition of entropy should be adopted, let us first examine the experimental evidence for residual entropy.
The residual entropy of glasses is obtained from calorimetric measurements of the specific heat during cooling from the liquid state \cite{Simon30,Parks28,Parks34,Bestul65}. Figure \ref{fig:CSofGlass} shows the measured specific heat and the derived entropy of a glass substance as functions of temperature.
The entropy of the glass, $S_{\rm gl}$, at $T=0$ is obtained by integrating the measured specific heat $C$ as,
\begin{equation}
S_{\rm gl}(0) =  \int_{T_{1}}^{0} \frac{C(T) }{T} dT 
  +S_{\rm liq}(T_{1}),
\label{eq:DeltaS-b}
\end{equation}
where $T_{1}$ is a reference temperature in the liquid phase. The entropy of the liquid, $S_{\rm liq}(T)$, is obtained by integrating the specific heat together with the latent heat measured during heating from the crystalline phase, which is the well-established method and there is no problem. The obtained value $S_{\rm gl}(0)$ consistently show that $S_{\rm gl}(0)>0$. In this way, the existence of residual entropy $S_{\rm res} = S_{\rm c} > 0$ is the well-established experimental fact. 
At this point, the experimental observation appears to contradict the Boltzmann interpretation of entropy: despite that the original $C-T$ curve is measured for a single glass sample, the resulting entropy includes the configuration entropy $S_{\rm c}$, which reflects the configurations of many possible samples. How can an individual sample ``know" the configurations of other samples?
This question of causality was the original question raised by Kivelson and Reiss \cite{Kivelson99}. Although it has been the subject of extensive debate
\cite{Speedy99,Moller06,Mauro07,Gupta07,Goldstein08,Reiss09,Gupta09,Aji10}, the original question has not yet been answered. 

Gibbs entropy provides a natural explanation for residual entropy. However, it introduces a different kind of conceptual difficulty. When Eastman and Milner first demonstrated the residual entropy of a random alloy, they explained it as follows: although a particular alloy sample adopts only one configuration, we do not know which configuration has been realized among the many possible ones \cite{Eastman33}. At first sight, this explanation appears convincing. However, it also gives entropy a subjective character, since its value seems to depend on our prior knowledge. Indeed, this viewpoint naturally raises a further question (see, for example, \cite{Sethna}, p.~86): if the exact atomic structure of a particular sample were completely determined, would the residual entropy $S_{\rm res}$ vanish? Whether entropy depends on our knowledge has been a central contention in the long-standing debate between information theory and thermodynamics \cite{Leff-Rex2,Jaynes79,Denbigh81}.

\subsection{Anthropomorphic nature of entropy}
\label{sec:Anthropomorphism}
The last issue of the dependence on prior knowledge is known as the {\it anthropomorphic} nature of entropy \cite{Jaynes65,Grad61}, an idea that was first proposed by Wigner.
For example, when the crystallographic structure of a paramagnetic rock-salt is studied, its entropy vanishes at $T=0$. However, if its magnetic properties are investigated, we find it of a nonzero value. In this way, the value of entropy changes depending on the information available to the observer. This observation naturally leads to the conclusion that entropy is not an intrinsic property of the material \cite{Wigner63}.
The central objective of information theory is to find the best statistical inference about the properties of the system, given a limited amount of our knowledge about the system \cite{Jaynes78}. The probability distribution is to be updated whenever new observations become available. Because information theory treats information in a very general sense, it is not constrained by energetic consideration. In thermodynamics, however, entropy is fundamentally related to energy through Eq.~(\ref{eq:ST-T}), and such freedom is not allowed.

The calorimetric method can determine only entropy differences through Eq.~(\ref{eq:defS}). The absolute entropy is fixed by the third law of thermodynamics. 
All materials share a common zero of entropy. This does not mean that the choice of the entropy origin is merely one convention among many possible alternatives. There are fundamental reasons why the entropy origin must be chosen in this way.
Consider a chemical reaction,
\begin{equation}
A + B \rightarrow AB.
\label{eq:chem-reaction}
\end{equation}
The equilibrium constant $K_{\rm eq} = [AB]/[A][B]$ is determined by the difference in the standard free energy $G^{0}$ between the reactants and products,
\begin{equation}
K_{\rm eq} = \exp \left( -\frac{\Delta G^{0}}{k_{\rm B}T} \right).
\label{eq:K-ChemEQ}
\end{equation}
If the entropy origins of different chemical species were chosen independently, the resulting equilibrium constant, $K_{\rm eq}$ would differ from the standard thermodynamic data, which are based on the third law, namely, $S(0)=0$ for all chemical species (\cite{Wilks}, \S 9.2). Likewise, for a phase transition $\alpha \leftrightarrow \beta$, if the two phases $\alpha$ and $\beta$ had different entropy origins, the predicted transition temperature would differ from the well-established experimental value. For ordinal crystals, since the third law holds well, neither equilibrium constants nor the transition temperatures are affected by whether magnetic measurements are performed, as in the case of paramagnetic rock salts. This observation indicates that the entropy origin remains unchanged even when additional variables are introduced.
Therefore, we need to formulate the third law in order to meet this property of entropy origin.

\subsection{Orientation of study}
\label{sec:Orientation}
Up to this point, we have encountered several deep difficulties in thermodynamics and statistical mechanics: the individual-or-species issue, the identification of an appropriate set of state variables, the origin of entropy, and the anthropomorphic character of entropy. The common feature underlying all of these difficulties is an ambiguity in the notion of {\it state} in thermodynamics.
In Introduction, we asked a series of questions of the form `What is X?", beginning with ``What is life?" These questions ultimately converge to a single question: ``What is a state?" Here, the term state refers to a thermodynamic state, not to a quantum state. This seemingly apparent question is the most difficult question. A thermodynamics state means an equilibrium state. Equilibrium is characterized by the absence of temporal change in state variables.
This immediately raises a further question: what is a state variable? State variables are defined only for equilibrium states, while equilibrium itself is specified through the constancy of the state variables. In this manner, we are led to circular argument \cite{Beauragard-note}. 
This fundamental dilemma has long plagued thermodynamicists. Until recently, this problem was thought to be unavoidable. A breakthrough was achieved by Gyftopoulos and Beretta (GB), who found a definition of equilibrium that does not rely on state variables \cite{Gyftopoulos}. Because the definition of equilibrium is the foundation of thermodynamics, this development has far-reaching consequences for many areas of thermodynamics theory, including its application of biology.

Therefore, the strategy of the present study is as following. First, we explain how the apparent dilemma between equilibrium and state variable is resolved (Sec.~\ref{sec:TDofSolids}). On the basis of unambiguous definition of state variables, the entropy of a thermodynamic state can be uniquely determined. We then show that the information of a material is distinct from its entropy and is instead conveyed by its state variables (Sec.~\ref{sec:Info}). Likewise, the notions of order and complexity can also be related to state variables rather than to entropy. 
Second, we show that solids possess hidden state variables, called {\it frozen} state variables, which underlie the apparent anthropomorphic nature of entropy (in Sec.~\ref{sec:FC}). A proper understanding of frozen state variables and configuration entropy is essential for resolving many of the apparent contradictions discussed above. 
Third, we emphasize that the set of state variables changes when a system undergoes a phase transformation. Consequently, the entropy significantly changes associated with appearance/disappearance of state variables, which requires careful treatment of the thermodynamic state space. On this basis, the existence of residual entropy can be reconciled with the third law. This framework also resolves the apparent contradictions between Boltzmann and Gibbs entropies that arises in many mixing phenomena (Sec.~\ref{sec:ConfS}).

At the end of this section, it is worthwhile to briefly mention the equivalence between the outcomes of thermodynamics and statistical mechanics. Although thermodynamics and statistical mechanics start from entirely different viewpoints---the former from macroscopic equilibrium and the latter from microscopic states---they ultimately lead to identical macroscopic predictions.
This point is important because it is often believed that statistical mechanics is more fundamental than thermodynamics. Such a belief may become an obstacle to resolving the conceptual contradictions discussed above. However, this brief is not correct. More than sixty years ago, Tisza and Mandelbrot demonstrated that thermodynamics and statistical mechanics are equivalent in the sense that they yield the same macroscopic prediction \cite{Tisza63,Mandelbrot64}. 

For example, consider a macroscopic quantity of specific heat ($C$) of a solid. Today, the standard procedure is first to calculate the phonon density of states, $g(u)$ ($u$ is the phonon energy), of the solid. This is a microscopic quantity.
From $g(u)$, the partition function ${\cal Q}$ is obtained as
\begin{equation}
{\cal Q}(\beta) = \int_{0}^{\infty} e^{-\beta u} g(u) du,
\label{eq:Zofphonon}
\end{equation}
where $\beta$ is inverse temperature. The partition function ${\cal Q}(T)$ then determines the free energy $F(T)$, from which the specific heat $C(T)$ is obtained.
Conversely, we may proceed in the reverse direction. By measuring the complete temperature dependence of the specific heat $C(T)$, we obtain $F(T)$ and ${\cal Q}(T)$. Now, by inverting the relation (\ref{eq:Zofphonon})---which is, in fact, an inverse Laplace transformation, the function $g(u)$ can be recovered \cite{Tisza63,Mandelbrot64}. 
\begin{equation}
g(u) = \int_{0}^{\infty} e^{\beta u} {\cal Q}(\beta) d\beta,
\label{eq:invLaplace}
\end{equation}
Thus, the Laplace transformation establishes a symmetric correspondence between the microscopic quantity $g(u)$ and the macroscopic quantity $C(T)$. The essential difference between thermodynamics and statistical mechanics therefore lies not in their predictions but in their interpretation. In thermodynamics, the physical meaning of the function $g(u)$ need not be specified; it serves only as an internal parameter that reproduces the observed macroscopic properties. This interpretation further implies that thermodynamic predictions remain valid even without detailed microscopic knowledge of $g(u)$.

\section{Thermodynamics of materials}
\label{sec:TDofSolids}

In Secs.~\ref{sec:TDofSolids} and \ref{sec:FC}, the framework of thermodynamics for materials is briefly presented in light of recent developments in thermodynamics theory. Although thermodynamics principles are the same as those presented in standard textbooks, the fact that the energy barriers in real materials are always finite values introduces time dependence into their responses. Since time dependence is generally not treated in conventional thermodynamics textbooks, special attention must be paid to the relevant timescale when real materials are investigated. 
This part is a summary of a previous study \cite{Shirai22-res}. All the enumerated definitions, corollaries, and postulates are the same as in \cite{Shirai22-res} with the same numbering, so that readers can readily refer to the original paper for more detailed explanations when necessary. The only exception is the terminology. In \cite{Shirai22-res}, the term {\it thermodynamic coordinate} was used in place of state variable. In the present paper, the latter term is retained to maintain consistency with the rest of the paper. 

\subsection{Equilibria of a solid}
\label{sec:EQ}
The approach of thermodynamics is to describe the properties of a system by focussing on its equilibrium states. Accordingly, the theory must begin with the definition of equilibrium. Unfortunately, as discussed in Sec.~\ref{sec:Orientation}, the conventional notions of equilibrium and state variables involve a circular argument. This difficulty was resolved by GB \cite{Gyftopoulos}, who succeeded in defining thermodynamic equilibrium without reference to state variables.
Suppose that a system is adiabatically connected to a weight, which is the only external device acting on the system.

\noindent
{\bf Definition 1: (Thermodynamic equilibrium)}
{\em It is impossible to change the stable equilibrium state of a system to any other state with its sole effect on the environment being a raise of the weight}. ([GB], p.~58)

\noindent
The expression ``with its sole effect on the environment being a raise of the weight" means, in more familiar terms, that work is extracted from the system without leaving any other effect on the surroundings. Definition 1 is regarded as a generalization of Clausius's statement of the second law, which asserts that it is impossible to extract work from a single heat bath. In the GB formulation, the single heat bath is replaced by an equilibrium state. To see the connection, suppose that work could be extracted from a single heat bath. Since no other source of energy is available, the extracted work must be supplied by a decrease in the internal energy of the bath. The decrease in the energy necessarily corresponds to cooling of the bath. Thus, work would have been extracted solely by cooling a single bath. This contradicts with Clausius's statement.

In Definition 1, the condition of ``its sole effect on the environment" must be stressed. This says that all constraints on the system are held fixed. For example, the volume of a gas in a container is held fixed by the walls of the container. If the constraint is removed, the gas expands and work can be extracted from it. In this case, the role of wall as the constraint is clear. In materials, however, the nature of the constraints is much less transparent.
The concept of constraints in materials was introduced more than a century ago by Gibbs \cite{Gibbs-TD}, who used the term {\it passive resistance}. His intention was to distinguish two categories in static states.
The first is a static state maintained by a balance of the active tendencies of the system. This corresponds to an ordinary chemical equilibrium, $A \leftrightarrow B$, in which the forward and backward reactions balance each other. 
The second is a static state in which no reaction occurs: $A \not \rightarrow B$. As an example, consider the synthesis of ammonia, 
\begin{equation}
{\rm N_{2} + 3 H_{2} \rightarrow 2 NH_{3}}. 
\label{eq:ammonia}
\end{equation}
According to standard thermochemical data, the Gibbs free-energy change of reaction (\ref{eq:ammonia}) at room temperature is $\Delta G^{0} = -16.2$ kJ/mol (Ref.~\cite{Yazawa-collection11}, p.~92). The negative sign in $\Delta G^{0}$ indicates that the reaction (\ref{eq:ammonia}) must spontaneously occur in the forward direction. In reality, however, no reaction occurs merely by mixing these two gases. Some obstacle must prevent the reaction from occurring. 

The obstacle was referred to by Gibbs as {\it passive resistance} and is now commonly described as a {\it constraint} \cite{Hatsopoulos}. Physically, a constraint originates from an energy barrier that inhibits a process. Even the macroscopic wall of a gas container can be regarded as a collection of microscopic energy barriers associated with the atoms constituting the wall.

\noindent
{\bf Definition 2: (Constraint)}
{\em A constraint $\xi_{j}$ is a means of restricting the range over which a particular variable $X(t)$ can vary. }

\noindent
A constraint may be expressed either by a hypersurface $\xi_{j}(X(t))=0$ or by an allowed region $\xi_{j}(X(t))>0$. 
For a solid, each atom cannot move beyond the unit cell to which it belongs. The boundary, $\xi_{j}$, of the $j$-th unit cell for the $j$-th atom acts as the corresponding constraint. Physically, the cell boundary is realized by the energy barrier $E_{b,j}$ surrounding the $j$-th atom. A crystal retains its structure through the set of constraints to which all atoms are subjected. In this context, lattice periodicity is not essential. Therefore, the notion of constraint applies equally well to nonperiodic materials, such as glasses. In this sense, the structure itself may be regarded as a collection of constraints that distinguishes one solid from another. Consequently, a solid possesses as many constraints, $\{ \xi_{j} \}$, as it has atoms, whereas a gas has only one constraint---the volume of a container.

Recognizing that the physical origin of a constraint is an energy barrier immediately leads to an important property of equilibrium: every equilibrium state has a finite lifetime. Suppose that a motion of the $j$-th type is hindered by an energy barrier $E_{b,j}$. The corresponding constraint remains effective only over a finite timescale $\tau_{j}$. This characteristic timescale $\tau_{j}$ of the equilibrium is determined by the energy barrier $E_{b,j}$ according to 
\begin{equation}
1/\tau_{j} = \nu_{j} e^{-E_{b,j}/k_{\rm B}T},
\label{eq:tau}
\end{equation}
where $\nu_{j}$ is the appropriate attempt frequency for the transition. This timescale $\tau_{j}$ is referred to as the {\it relaxation time}. Because all energy barriers are finite, no equilibrium state can exist indefinitely. 
For example, once the gas molecules become uniformly distributed within a container, the gas is regarded as being in equilibrium. However, even if the container is made of a robust metal, a high-pressure gas cannot be confined indefinitely, because of unavoidable leakage over sufficiently long times. Likewise, all crystals eventually undergo degradation through chemical reaction with the environment, impurity diffusion, the formation of dislocations, and other irreversible processes.

Equilibrium states can therefore be identified only with respect to the relevant timescale. This recognition is important because, in glass physics, many researchers regard the present state of a glass as a nonequilibrium states on the basis of the observed aging phenomena. However, aging is common to all materials. Even a gas that is uniformly distributed within a container undergoes gradual changes until the pressures inside and outside become equal. From the present perspective, the current state of a glass is already an equilibrium state. Readers who adhere to the conventional view that glasses are intrinsically nonequilibrium states are referred to the review in Ref.~\cite{Shirai-SH-Liquids25} (\S 5.2 and 5.3) and the study on hysteresis in Ref.~\cite{Shirai26-hysteresis}.

\noindent
{\bf Corollary 1: (Equilibrium states of a solid)}
{\em If the structure of a solid remains unchanged during a time period $\tau$, the solid is in an equilibrium state over that time interval, regardless of whether its structure contains defects.}

Corollary 1 implies that a solid possesses many equilibrium states for a given $T$ and $V$. This is the point most significantly deviated from the conventional view. Traditionally, it has been assumed that, for a given $T$ and $V$, a solid has only one equilibrium state, namely, the lowest-energy state. This is incorrect. Thermodynamically, an equilibrium state is the state of minimum free energy, which includes the entropic contribution. Consequently, a crystal containing the equilibrium concentration of defects has a lower free energy than the perfect state. 
By creating defects, a crystal can possess numerous defect structures that remain unchanged over the relevant timescale. A schematic potential-energy landscape of a solid is shown in Fig.~\ref{fig:basins}. According to Definition 1, these defect states are equilibrium states because no work can be extracted from them without producing an effect on the environment. A more detailed discussion of this point is given in \cite{Shirai26-hysteresis}, in particular in Sec.~3.

From Eq.~(\ref{eq:tau}), we see that the equilibrium timescale depends strongly on temperature. As $T$ approaches 0, any finite energy barrier becomes infinitely large relative to the thermal energy $k_{\rm B}T$. It therefore follows that

\noindent
{\bf Corollary 2: (Equilibrium at $T=0$)}
{\em As the temperature approaches zero, every existing solids retain its equilibrium state for an effectively infinite time, unless the environment is altered.}

\noindent
Since all relaxation times diverges as $T \to 0$, the distinction among stable, metastable, and frozen states disappears. 

\subsection{State variables and the fundamental relation of equilibrium}
\label{sec:SV}

\begin{figure}[htbp]
    \centering
    \includegraphics[width=140 mm, bb=0 0 540 320]{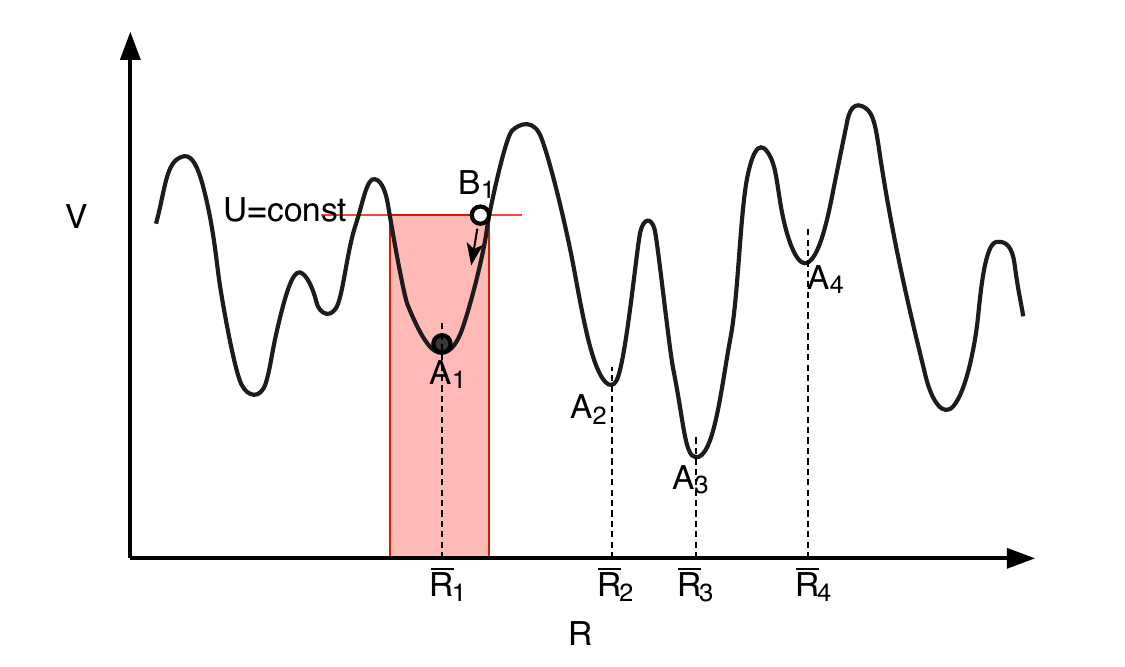} 
  \caption{Schematic potential-energy landscape of a solid. The equilibrium states $A_{j}$ are specified by state variables $\{ \bar{\bf R}_{j} \}$. They form discrete points in the space of atomic coordinates, each corresponding to a configuration $K$. Starting from state $B_{1}$, for a given $U$, the solid can access only the microscopic states within the red region.
  } \label{fig:basins}
\end{figure}

The common feature of every state variable $X$ is that it remains constant in time at equilibrium. From microscopic viewpoint, however, every property $X$ of a material exhibits thermal fluctuations at finite temperatures, giving rise to a small time dependence $X(t)$, as $X(t) = \bar{X} + \Delta X(t)$, where $\bar{X}$ is the mean value of $X(t)$ and $\Delta X(t)$ is the deviation from the mean value. Even the boundary of a gas confined in a container fluctuates on the atomic scale. Such microscopic fluctuations are irrelevant to the internal energy $U$ of the system. Only the time-averaged value, $\bar{X}$, is relevant to $U$, as indicated by Eq.~(\ref{eq:FRE-gas}) for the gas case. Thus, the volume $V$ appearing in Eq.~(\ref{eq:FRE-gas}) is, in fact, the time average of the microscopically fluctuating quantity $V(t)$. 
Similarly, the density of a gas contains a time-dependent microscopic component, $\rho(t) = \bar{\rho} + \Delta \rho(t)$. Only the time averaged part $\bar{\rho}$ is relevant to its equilibrium properties of the gas and therefore serves as a state variable.
It is therefore natural to define a state variable as the time average of a dynamical variable $X(t)$, as suggested by Callen (\cite{Callen}, p.~5).

\noindent
{\bf Definition 3: (State variable)}
{\em A state variable is the time-averaged value of a dynamical variable under a constraint that restricts the range within which the instantaneous value of the dynamical variable can vary}.

In mathematical form, a state variable, $X_{j}$, is written as
\begin{equation}
X_{j} = \frac{1}{t_{0}} \int_{\xi_{j}} X(t) dt,
\label{eq:t-aveX}
\end{equation}
where $t_{0}$ is the time interval over which the average is taken. $t_{0}$ may be chosen arbitrarily, provided that it is shorter than the relaxation time $\tau_{j}$. In Definition 3, it is assumed that the time average converges to a well-defined value; namely, the averaged value is independent of the sampling time $t_{0}$. 
The time-averaged position $\bar{x}$ of an atom in a gas is indeterminate, and hence $\bar{x}$ cannot be a state variable. In contrast, in solids, the position of atom $j$ fluctuates about its equilibrium position $\bar{\bf R}_{j}$, as ${\bf R}_{j}(t) = \bar{\bf R}_{j} + {\bf u}_{j}(t)$, where ${\bf u}_{j}$ is the displacement from $\bar{\bf R}_{j}$. Therefore, $\bar{\bf R}_{j}$ qualifies as a state variable. This conclusion may be the greatest departure from the traditional understanding of thermodynamics. However, this conclusion correctly reflects experimental facts. 
Suppose that an atom in a crystal is displaced, for example, by electron irradiation. The displaced atom $j$ eventually settles into a new equilibrium position, ${\bf R}'_{j}$, at one of the interstitial sites. This defect configuration is a new equilibrium state because no further change in the atom position occurs. It also has a different energy from that of the unperturbed crystal. Such an energy change cannot be expressed by a formula that involves only $V$, as in Eq.~(\ref{eq:FRE-gas}). Clearly, for solids, $U$ depends on the atomic structure and therefore must be a function of $\{ \bar{\bf R}_{j} \}$. This is the most fundamental difference from the gas state, for which there is no internal structure and thus $U$ does not depend on the atomic positions. Therefore, we conclude:

\noindent
{\bf Corollary 3: (State variables of a solid)}
{\em The equilibrium position of each atom $\bar{\bf R}_{j}$ is a state variable of a solid.}

\noindent
This is not to say that the instantaneous position ${\bf R}_{j}$ of an atom is a state variable. Only the {\it discrete} equilibrium positions $\bar{\bf R}_{j}$ qualify as state variables. Thus, a state variable $\bar{\bf R}_{j}$ is not a continuous variable. This is an important difference from continuous state variable such as volume. 

As already stated by Corollary 1, a solid possesses many equilibrium states for a given $T$ and $V$. Corollary 3 provides the quantitative basis for this multiplicity of equilibrium states. 
This property has been demonstrated in studies of glass transition \cite{Shirai20-GlassState,Shirai22-SH,Shirai23-Silica,Shirai-SH-Liquids25} and hysteresis phenomena in solids \cite{Shirai26-hysteresis}.
Furthermore, it has a significant implications for biology. One of fundamental questions in biology is the apparent dilemma between the stability of genetic system and the diversity of evolution (more specifically, phenotypic plasticity)\cite{Kaneko09,Kauffman95}. Although the preservation of a species through DNA appears to be highly robust, it seems paradoxical that a great diversity of species and individuals nevertheless emerges. From the present perspective, these two properties are not mutually exclusive but rather represent different consequences of the multiplicity of equilibrium states in a solid. This point is further elaborated in following sections.

With the above definitions of equilibrium and constraint, the second law of thermodynamics can be formulated as follows:

\noindent
{\bf Postulate 1: (The second law of thermodynamics)} 
{\em Among all the states of a system that have a given $U$ and are compatible with the given constraints $\{ \xi_{j} \}$, there exists one and only one stable equilibrium state.}  [GB, p.~63]

\noindent
The significance of Postulate 1 can be understood with the aid of the potential-energy landscape shown in Fig.~\ref{fig:basins}. Let us choose an arbitrary state $B_{1}$---generally a nonequilibrium state---as the initial state of a system with the energy $U$. The initial positions $\{ {\bf R}_{j}(0) \}$ determine the potential basin in which the atoms are confined, thereby defining the constraints $\{ \xi_{j} \}$. Because $U$ is fixed, the system can access only those states $\{ {\bf R}_{j}(t) \}$ within the shaded region of the figure. The system then evolves spontaneously and eventually reaches the equilibrium state $A_{1}$, corresponding to the equilibrium configuration $\{ \bar{\bf R}_{j}^{1} \}$. Within the given constraints $\{ \xi_{j} \}$, this equilibrium state is unique. It is also the state of maximum entropy. Thus, Postulate 1 recovers the familiar maximum-entropy principle for an isolated system. Equivalently, for a fixed entropy, the equilibrium state corresponds to the minimum of the appropriate thermodynamic potential: see \cite{Gibbs-TD} (p.~56). 

However, significance of Postulate 1 is not limited to this point. If the system starts from a different initial state $B_{2}$, located in another potential basin, it evolves toward a different equilibrium state $A_{2}$. The state $A_{2}$ is the maximum-entropy state within the constraints defining that basin. As shown in Fig.~\ref{fig:basins}, there are many basins, and hence many such maximum-entropy states exist. 
Thus, the term ``maximum entropy'' has meaning only with respect to the accessible region of the state space determined by the constraints. The maximum is therefore local rather than global. Within each constrained region, the equilibrium state is unique and corresponds to the extremum of the appropriate thermodynamic potential. In this local sense, the maximum-entropy principle and the minimum-potential principle are equivalent descriptions of equilibrium.

The above consideration shows that there is a one-to-one correspondence between a set of constraints $\{ \xi_{j} \}$ and the resulting equilibrium state. Independently, Reiss established a one-to-one correspondence between constraints and state variables \cite{Reiss}.
Together, these results establish a one-to-one correspondence between equilibrium states and state variables.
It follows that the equilibrium state of a system is completely and uniquely specified by the set $\{ X_{j} \}$. Conversely, for a given set of constraints $\xi_{j}$, each equilibrium state uniquely determines the corresponding values of the state variables $X_{j}$. 

\noindent
{\bf Corollary 4: (Existence of the fundamental relation of equilibrium)}
{\em For a system with a fixed $U$ under a given set of constraints, the state variables $\{ X_{j} \}$ are uniquely determined at equilibrium.}

\noindent
The statement in Postulate 1 that there exists one and only one equilibrium state for a given set of constraints implies the existence of a state function that uniquely characterizes that equilibrium. This state function is the entropy $S$, which is exactly the starting assumption of the axiomatic approach of thermodynamics (Sec.~\ref{sec:thermodynamic-entropy}). Accordingly, there exists a functional relationship between $S$ and the complete set of state variables,
\begin{equation}
S=S(U, X_{1}, \dots, X_{M}).
\label{eq:FundamentalE}
\end{equation}
This expression is fully consistent with the axiomatic definition of entropy, Eq.~(\ref{eq:Sfunction}).
This equation is another form of the FRE in entropy expression, Eq.~(\ref{eq:Sfunction}). 
It is worth noting that ``the FRE is derived as a rigorous consequence of the first and second laws, not as a consequence either of difficulties related to exact calculations and lack of knowledge, or a need to describe complicated physical problems by a few gross macroscopic averages" (\cite{Gyftopoulos}, p.~119).

The FRE has the following important property \cite{Gibbs-TD, Callen, Gyftopoulos}.

\noindent
{\bf Corollary 5: (Completeness of state variables)}
{\em Any thermodynamic property of a system can be derived from the fundamental relation of equilibrium.}

\noindent
If the set of state variables were not complete, the one-to-one correspondence between equilibrium states and state variables would not hold.
The $M$ state variables span an $M$-dimensional space. This space is called the {\em thermodynamic state space}. The adjective ``thermodynamics" is used because the term {\it state space} is already widely used in quantum mechanics with a different meaning.

\noindent
{\bf Definition 4: (Thermodynamic state space)}
{\it The thermodynamic state space ${\mathscr A}$ spanned by the complete set of $M$ state variables, $\{ X_{j} \}$, is expressed as
\begin{equation}
{\mathscr A}=(U, X_{1}, \dots, X_{M}).
\label{eq:TDspace}
\end{equation}
}
\noindent
For convenience, the internal energy $U$ is treated as the zeroth coordinate and is not counted in the dimensionality of the thermodynamic state space. 
Equation (\ref{eq:TDspace}) indicates that the value of entropy depends on the thermodynamic state space ${\mathscr A}$ on which $S$ is expressed:
\begin{equation}
S^{\mathscr A}=S(U, X_{1}, \dots, X_{M}).
\label{eq:SinA}
\end{equation}
When the same physical state is represented in different thermodynamic spaces ${\mathscr A}$ and ${\mathscr B}$, one generally finds that $S^{\mathscr A}(A) \neq S^{\mathscr B}(A)$, even though the two values refer to the same physical state. As will be discussed later, this observation provides the basis for explaining the anthropomorphic character of entropy.

Suppose that the entropy of state $A$ is evaluated in a thermodynamic state space ${\mathscr A}$, giving $S^{\mathscr A}(A)$, while the entropy of another state $B$ is evaluated in a different thermodynamic state space ${\mathscr B}$, giving $S^{\mathscr B}(B)$.

\noindent
{\bf Corollary 6: (Comparability condition)}
{\em A necessary condition for comparing two entropy values, $S^{\mathscr A}(A)$ and $S^{\mathscr B}(B)$, is that the thermodynamic state spaces ${\mathscr A}$ and ${\mathscr B}$ have the same dimensionality.}

\noindent
When the two states, $A$ and $B$, are evaluated on the same space, it is always possible to connect them by a reversible path in Eq.~(\ref{eq:defS}), because the state can be changed continuously from $A$ to $B$ {\em without breaking constraints that define that thermodynamic state space}.

The term {\it configuration} is now used in a rigorous thermodynamic sense to refer exclusively to an equilibrium configuration. 

\noindent
{\bf Definition 5: (Configuration)}
{\em A configuration is specified by a complete set of equilibrium positions of atoms $\{ \bar{\bf R}_{j} \}$. } 

\noindent
Since the atoms in a liquid do not possess equilibrium positions, the term configuration is not used for liquids in the present framework. In this respect, readers may need care when comparing to other references. When there are $N_{\rm c}$ distinct configurations, the thermodynamic state space is expressed as:
\begin{equation}
{\mathscr A}=(U, \{ \bar{\bf R}_{j}^{(K)} \}  ),
\label{eq:TDSDiscrete}
\end{equation}
where $(K)$ represents collectively all equilibrium configurations $K =1, \dots, N_{\rm c}$. The set $\{ \bar{\bf R}_{j}^{(K)} \}$ may be regarded as a coordinate transformation from the atomic index $j$ to configuration index $K$, analogous to the normal modes transformation from atomic displacements to normal-mode coordinates.

As noted above, the state variables $\bar{\bf R}^{K}$ are defined only at discrete points in the multidimensional space of atom positions $\{ {\bf R}_{j} \}$. These points are separated by energy barriers. A transition from one configuration $K$ to another $K'$ requires removal or overcoming of the corresponding constraints and is therefore generally irreversible unless special precautions are taken. Consequently, the entropy change associated with such a transition cannot, in general, be evaluated by the calorimetric method based on Eq.~(\ref{eq:defS}). Hysteresis is a typical example, because hysteresis is essentially a series of successive transformations between adjacent configurations $K \to K'$ and thus is irreversible (\cite{Shirai26-hysteresis}, \S~3.4).

\subsection{Degeneration of thermodynamics state space}
The dimensionality $M$ of the thermodynamic state space varies substantially among different phases.
In a transition from a solid to a gas, $M$ is reduced from $3 N_{\rm at}$ to just one:
\begin{equation}
( \bar{\mathbf R}_{1}, \dots, \bar{\mathbf R}_{N_{\rm at}}) \; \rightarrow V.
\label{eq:ri-ni}
\end{equation}
This reduction is referred to as the {\it degeneration} of the thermodynamic state space \cite{Shirai18-StateVariable}. The reverse process is called the {\it generation} of the thermodynamic state space. Note that this usage of degeneration, which is borrowed from mathematics, is different from usual one in physics \cite{Shirai18-StateVariable}.
The degeneration is accompanied by an increase in entropy, because the quantities that were previously well-defined state variables $\bar{X}_{j}$ no longer posses definite time-averaged value. Consequently, they ceases to function as state variables in the thermodynamic description. 
If the dimensionality $M$ is interpreted as representing the information content of a material---a viewpoint discussed in Sec.~\ref{sec:Info}---the degeneration of the thermodynamic state space may be regarded as a loss of information. This interpretation resembles the notion of missing information in information theory. The crucial difference, however, is that the missing information is an intrinsic property of the material rather than a consequence of an observer's lack of knowledge \cite{Denbigh81}. More precisely, what is lost in the degeneration process is the set of available state variables, and the accompanying entropy increase is therefore a consequence of reduction in the number of state variables. 

Corollary 6 states that, in general, comparing entropy values between different phases requires the two thermodynamic state spaces to have the same dimensionality. 
In the vaporization of a solid, state variables $\bar{\bf R}_{j}$ cease to exist because the corresponding constraints $\xi_{j}$ are removed. Since removing constraints is, in general, an irreversible process, the calorimetric method based on Eq.~(\ref{eq:defS}) cannot be applied to evaluate the associated entropy change. 
However, the phase transition can be carried out reversibly by bringing the two phases into equilibrium through heat exchange at the equilibrium condition. In this situation, the entropy increase caused by the change in the thermodynamic state space is exactly compensated by the entropy flow associated with the latent heat of transition.
For example, at the melting temperature $T_{m}$, the entropy change $\Delta S_{\rm tr}$ of transition is related to the latent heat $\Delta H_{\rm tr}$ by $\Delta S_{\rm tr} = \Delta H_{\rm tr}/T_{m}$, so that the entropy difference between the solid and liquid phases can be determined by calorimetric measurement as is expected.

\section{Frozen state variables}
\label{sec:FC}
In general, the time average of an atom position in a gas is indeterminate, and thus $\bar{\bf R}_{j}$ cannot serve as a state variable of a gas. However, strictly speaking, there is one notable exception. When a gas is confined within a fixed container, the time-averaged position $\bar{\bf R}_{j}$ of every atoms is the center of mass, ${\bf R}_{\rm cm}$. This value is identical for all atoms and is trivial from the viewpoint of the thermodynamics state. Although ${\bf R}_{\rm cm}$ satisfies the definition of state variable (Definition 3), the internal energy $U$ of the gas is independent of it.

\noindent
{\bf Definition 6: (Frozen state variable)}
{\em A frozen state variable is a state variables that exists but does not affect the fundamental relation of equilibrium to the thermodynamic problem under consideration.} 

Only those state variables on which the FRE depends are relevant to the thermodynamic description. The properties of water in a container do not vary when measured at different elevations $H$ at which the container is held; they depend only on $T$ and $V$. The FRE is expressed by these two variables, $(T, V) \equiv \{ X_{j}^{0}\}$. However, there are cases in which $H$ does affect the properties of water. In a hydraulic power plant, the elevation $H$ is the most important variable, and the FRE depends on $H$ \cite{Kline57,Hatsopoulos}. If the gate of the upper reservoir is closed so that no water can flow between the upper and lower water, $H$ no longer enters the FRE. In this case, $H$ is a frozen state variable.
Let us denote frozen state variable by $\hat{X}$ and let us take this variable into the FRE. The FRE of the water in the upper reservoir is expressed as $S(\{ X_{j}^{0}\}; \hat{H}) = S_{X}(\{ X_{j}^{0} \}) + S_{H}(\hat{H})$. The part $S_{H}(\hat{H})$ is constant as long as the gate is closed. The role of the constraint in this case is to completely inhibit $\hat{H}$ from varying. When the gate is opened (the constraint is removed), the frozen variable $H$ is activated to vary. 
The upper water then flows downward, which is a relaxation process toward a new equilibrium state. Once the relaxation is complete, $H$ gains a new constant value of the lower reservoir.

As illustrated in this example, a frozen state variable arises when its value is completely fixed by a constraint. Consequently, it does not vary with temperature under this constraint and is therefore thermodynamically inactive. 

\noindent
{\bf Definition 7: (Active state variable)}
{\em A state variable that is not frozen is an active state variable.}

\noindent
Any state variable appearing explicitly in the FRE is an active state variable, because variations of that variable influence the equilibrium properties of the system through the FRE. An active state variable introduces temperature dependence into its thermodynamic properties via Eq.~(\ref{eq:U-FREsolid}). 
For a solid, the equilibrium position $\bar{\bf R}_{j}$ is an active state variable. Its thermodynamic significance originates from the thermal fluctuations around the equilibrium position, expressed as  ${\bf R}_{j}(t) =\bar{\bf R}_{j} + {\bf u}_{j}(t)$, where ${\mathbf u}_j(t)$ is the vibrational displacement. Through this coupling, the entropy and other thermodynamic properties acquire temperature dependence.
 In contrast, a frozen state variable is fixed by a constraint and therefore does not contribute to thermodynamic responses. The apparent anthropomorphic nature of entropy discussed in Sec.~\ref{sec:Anthropomorphism} originates from the presence of frozen state variables \cite{Shirai22-res}.

\noindent
{\bf Corollary 7: (Extending thermodynamic state space)}
{\em It is possible to extend the thermodynamic state space ${\mathscr A}=(U, \{ X_{j}\} )$ by including frozen state variables $\{ \hat{Y_{j}}\}$, thereby forming ${\mathscr B}=(U, \{ X_{j}\}; \{ Y_{j}\} )$, as long as the constraints of the frozen state variables are fixed.}

\noindent
Under this extension, the entropy of state $A$ transforms as
\begin{equation}
S^{\mathscr B}(A) = S^{\mathscr A}(A)+S_{0},
\label{eq:SBeqSAplus0}
\end{equation}
where $S_{0} = S(\{ \hat{Y}_{j} \} )$ is a constant. 

Let us denote the thermodynamic state space spanned by the active state variables as ${\mathscr A} = (U, \{ \bar{\bf R}_{j}^{1} \})$.

\noindent
{\bf Definition 8: (Active configuration)}
{\em The active configuration consists solely of the active state variables $\{ \bar{\bf R}_{j}^{1} \}$. }

\noindent
All remaining configurations are treated as frozen configurations because they are fixed by constraints and do not participate in the thermodynamic description of the active equilibrium state.
The full thermodynamic state space is written as
\begin{equation}
{\mathscr B}=(U, \{ \bar{\bf R}_{j}^{1} \}; \{ \hat{\bf R}_{j}^{(K')} \} ),
\label{eq:TDS-FC1}
\end{equation}
where $\{ \hat{\bf R}_{j}^{(K')} \} = \{ \hat{\bf R}_{j}^{K'} \}_{K' = 2, \dots, N_{\rm c} }$.
At sufficiently low temperatures, only the configuration $K=1$ remains an active configuration, whereas all other configurations are frozen. Therefore, we can arrive at the third law of thermodynamics but with more careful statement:

\noindent
{\bf Postulate 2: (The third law of thermodynamics)}
{\em The entropy $S^{\mathscr A}$ of any configuration of a solid approaches zero as $T \to 0$, provided that the entropy is evaluated in the thermodynamic state space ${\mathscr A}$ containing only the active configuration, ${\mathscr A}=(U, \{ \bar{\bf R}_{j}^{1} \})$.} 

Because entropy is a state function (Corollary 4) and only the active configuration is the equilibrium state near $T=0$, only this way of assigning the entropy value at $T=0$ is legitimate. For the case of a glass, the configuration occupied by the sample is the active configuration $\{ \bar{\bf R}_{j}^{1} \}$ and other configurations ($K \neq 1$) are frozen configurations. Indeed, the low-temperature specific heat $C(T)$ of a glass is determined solely by the active configuration: other configurations make no contribution to the measured specific heat. Therefore, even for glasses, the entropy must vanish at $T=0$.

The experimentally observed residual entropy is not the thermodynamic property of the {\em present} glass state, meaning that the value is evaluated on the thermodynamic space of the high-temperature phase, ${\mathscr B}$. 
The pitfall of the calorimetric method based on the integration form of Eq.~(\ref{eq:DeltaS-b}) is that the entropy value is affected from the previous values if the integration path is not reversible. The current value $S_{\rm gl}(0)$ in Eq.~(\ref{eq:DeltaS-b}) involves the integration constant $S_{\rm gl}(0)$, which is the value evaluated in the thermodynamic state space of liquid, ${\mathscr B}$.
The residual entropy comes from this integration constant in Eq.~(\ref{eq:DeltaS-b}). It represents the entropy contribution associated with state variables that were previously active in the high-temperature thermodynamic state space ${\mathscr B}$ but become frozen in the low-temperature thermodynamic space ${\mathscr A}$.
For ordinal crystals, this problem does not arise. The generation and degeneration of the thermodynamic state space occur reversibly and are accompanied by the latent heat of the transition. Consequently, the calorimetric integration remains reversible and no residual entropy is produced. For further details of the third-law issue, refer to \cite{Shirai22-res}.

\section{Information of materials and the time invariance}
\label{sec:Info}

\subsection{Randomness does not imply lack of order}
\label{sec:Random}

Equilibrium implies the absence of time evolution. Accordingly, state variables remain constant in time, as expressed in Definition 3.
This constancy may be characterized in several equivalent ways: invariance under time averaging, invariance with respect to thermal fluctuations, and invariance with respect to the sampling-time of a measurement. For brevity, the term {\em time invariance} will be used in this paper. Readers should note, however, that this usage differs from the usual meaning of time invariance in dynamics, where it refers to the conservation in the microscopic level. In the present context, the time invariance means that repeated measurements of a property yield the same value as long as the system remains in equilibrium. As will be shown below, this notion of time invariance plays a central role in the interpretation of entropy and related thermodynamic quantities.

First, let us consider the relation between order and entropy, which was posed at the beginning of this paper. In physics, the concept of order has traditionally been associated with the presence of regularities in geometrical patterns \cite{Weyl52}. Symmetry provides the most familiar example of such regularities. Any defect disturbing crystallographic symmetry is regarded as a source of disorder, and glasses are therefore often considered to be fully disordered system. Despite this conventional view, calorimetric measurements reveal the existence of order parameters through the characteristic jump in the specific heat at the glass transition \cite{Davies53a,Nemilov-VitreousState}. The author's group has shown that these glass order parameters are found to be the equilibrium atom positions, $\bar{\bf R}_{j}$ \cite{Shirai22-SH,Shirai23-Silica}. This suggests that order parameters are fundamentally equivalent to state variables. Indeed, in a subsequent study, this equivalence was established explicitly \cite{Shirai25-OrderParams}.

\noindent
{\bf Corollary 8: (Order of materials)}
{\em The order in a material is represented by its state variables.}

This equivalence can be understood by considering magnetic materials. The most familiar example of an order parameter is the magnetization ${\bf M}$ of a ferromagnet. In a ferromagnetic state, the magnetic moments of all atoms align in the same direction, say the $z$ direction, so that ${\bf M} = N_{\rm at} {\bf m}$, where ${\bf m}$ is the local magnetic moment of an atom. In this case, a single order parameter ${\bf M}$ characterizes the magnetic order, and hence the number of order parameters is $N_{\rm op}=1$. There are many other magnetic orders, although the total magnetization ${\bf M}$ may vanish. Antiferromagnetic order is created by an alternate arrangement of two kinds of local moments, ${\bf m}$ and $-{\bf m}$. In this case, $N_{\rm op}=2$. If the magnetic unit cell contains three independent local moments, ${\bf m}_{1}$, ${\bf m}_{2}$ , and ${\bf m}_{3}$, it creates a ferrimagnetic order with $N_{\rm op}=3$. More generally, any linear combination, $\sum_{j}^{s} {\bf m}_{j} \cos (q_{j} x)$ ($q_{j}$: the wavevector of $j$-th moment), of $s$ independent local moments creates a ferri- or antiferro-magnetic order. In the limit $s \rightarrow \infty$, the system approaches a spin-glass state, for which $N_{\rm op} = N_{\rm at}$. In this limit, any spatial regularity is lost. Nevertheless, each atom retains a nonzero local moment: the time average $\overline{ {\bf m}_{j}(t)}$ has a constant value ${\bf m}_{j}$. This is the essential distinction from a paramagnetic state, in which the time average $\overline{ {\bf m}_{j}(t)}$ of a nonzero instantaneous moment vanishes. The above examples indicate that order should not be identified with spatial regularity. Rather, order is characterized by the existence of time-invariant quantities. Crystal symmetries significantly reduce the number of independent order parameters. In the most general case, however, the maximum number of the order parameters is $3 N_{\rm at}$, which coincides with the dimension of the thermodynamic state space of a solid.

The inadequacy of associating entropy with disorder has already pointed out by several authors \cite{Ben-Naim,Rosenkrantz83, Landsberg84,Grandy,Denbigh89,Styer00}. Despite this inadequacy, because no one succeeded to give the alternative definition of order and randomness on the thermodynamics grounds, the traditional interpretation has continued to prevail among researchers.
The above example demonstrates that spatial regularity is not essential for the order in solids. Rather, the time-average invariance is the genuine ingredient of the order. Spatial symmetries merely reduce the number of independent variables required to describe the ordered state.
From this viewpoint, it is not surprising to find order in glasses and biological systems despite their lack of apparent spatial regularity. 
Now, the true meaning of the notion ``aperiodic crystal" introduced by Schr\"{o}dinger can been understood on this ground. Nonperiodicity does not imply the absence of order. As long as a material possesses definite time-invariant state variables, it retains a definite structure. The amount of information is not determined by the degree of periodicity but by the number of independent state variables required to specify the structure. This observation motivates the following discussion.

\subsection{Information is contained in state variables}
\label{sec:Information}

Second, the notion of time invariance also provides a natural definition of the information contained in a material.
In physics, a material is identified through its measurable properties, such as those obtained by chemical analyses and the X-ray diffraction method. There are many properties $\{ Y_{k} \}$ in a material. According to Corollary 5, each property is uniquely determined by a set of state variables $\{ X_{j} \}$: 
\begin{equation}
Y_{k}=Y_{k}(U, X_{1}, \dots, X_{M}).
\label{eq:YUX}
\end{equation}
By converting variables in Eq.~(\ref{eq:YUX}), one can choose $M$ independent properties $\{ Y_{k} \}$ as the state variables. Therefore, the complete set of state variables contains all information necessary to determine the measurable properties of the material. In this sense, all information about a material is encoded in its state variables.
From this, we arrive at the following interpretation of the information contained in a material.

\noindent
{\bf Corollary 9: (Information of materials)}
{\em  The information of a material is carried by its state variables.}

\noindent
Because of the one-to-one correspondence between constraints and state variables (Postulate 1),
Corollary 9 can be restated as follows: constraints carry information \cite{Caticha21}. Entropy is not a quantity to represent the information contained in a material. Rather, as indicated by Eq.~(\ref{eq:FundamentalE}), entropy characterizes the magnitude of the fluctuations associated with the information encoded in the state variables $\{ X_{j} \}$. This distinction naturally resolves the long-standing dilemma between the information and entropy discussed in Sec.~\ref{sec:Dilemmas}. Information is represented by the state variables themselves, whereas entropy quantifies the fluctuations associated with those state variables.

The sampling-time invariance of state variables---and therefore of the order---ensures the preservation of information of a material. This principle of information can be extended beyond physics.
Consider a town map. The spatial arrangement of houses in a town generally does not have periodicity. However, each house is assigned a unique address. A mail carrier can deliver postal materials to any house once he has access to the town map. However, if residents move frequently, the mail carrier cannot deliver the postal materials correctly. In order to correctly deliver them, the mail carrier must ensure that the address of today is the same as yesterday. Thus, the essential requirement for information is not spatial regularity but temporal persistence. The record of addresses preserves the correspondence between residents and locations, thereby maintaining the order of the system.

An English text would appear to be merely a random sequence of letters to someone who does not understand English. However, as long as the word ``apple" is always used to refer to the object apple, this word conveys meaningful information. If everyone uses it differently, this word loses its sense. 
Information becomes meaningful only when the same answer is obtained whenever the same question is asked. This property may be called the sampling invariance.
A similar argument applies to DNA. If we do not know the meaning of nucleotides sequence, a DNA molecule appears to be merely a random arrangement of nucleotides. The sequence acquires meaning only because the arrangement is preserved over time. If the nucleotide sequence changed continuously, the information encoded in DNA could not be transmitted reliably, and the reproduction of living organisms would become impossible.

In telecommunication theory, the primary concern is how efficiently information can be transmitted and how compactly data can be encoded \cite{Shannon48,Caves93,Zurek99,Cover-Thomas}. The statistical structure of a message plays a central role in this context. In particular, the probability distribution $p_i$ of an event $i$ provides the basis for analyzing communication efficiency, and the information-theoretic entropy, Eq.~(\ref{eq:HInform}), serves as a fundamental quantity.
From the thermodynamic viewpoint, however, the meaning of information differs from that employed in information theory. As illustrated by the language example above, the significance of the word ``apple'' cannot be inferred solely from the statistical regularity of the sequence of letters. The word conveys information only because a stable correspondence has been established between the word and the object it denotes. Outside the linguistic community that shares this convention, the word loses its meaning. It is the set of social rules maintaining this correspondence that makes communication possible.

Similarly, the sequence of nucleobases in DNA plays a role analogous to that of language. From this viewpoint, seeking statistical regularities in the nucleotide sequence may not be the essential issue. Rather, the existence of a definite sequence itself carries information. A biological word, called a codon, consists of three nucleobases. Each codon specifies a particular amino acid, and the resulting sequence of amino acids determines the structure of a protein. For example, the codon ``GUC'' codes valine.
There appears to be no direct chemical necessity for the particular correspondence between codons and amino acids \cite{Monod,Smith97}. Rather, the correspondence functions as a rule. This rule is mediated by tRNA, where the site connecting to mRNA is spatially well separated from the site connecting to the amino acid \cite{Life-9} (Sec.~9.4). The specific codon--amino-acid correspondence may have been selected contingently from many possible alternatives during early biological evolution \cite{Monod,Thaxton20,Kauffman95,Murphy-Oneill97}. Once this particular correspondence was selected, mechanisms for preserving this correspondence evolved and have maintained it throughout subsequent biological history.
The essential point from the present perspective is that the correspondence remains invariant over time. It is this time invariance of the rule that enables the storage and transmission of biological information.
One of the deepest questions in the life sciences is how such a template-based mechanism originated in a prebiotic environment \cite{Thaxton20}. Although this question may never be answered completely, the argument presented in Appendix A may provide a new perspective and advance our understanding by one step.


\subsection{Complexity and information}
\label{sec:Complexity}
Finally, the capability of a material to store information is discussed.
As described in Introduction, this capability is often associated with the notion of complexity, which leads to the dilemma between the complexity and entropy. As clarified above, however, the information contained in a material is distinct from its entropy. Therefore, it is inappropriate to characterize the information-storage capability of a material in terms of its entropy. 
Instead, the present study proposes that complexity should quantified by the dimension $M$ of the thermodynamic state space.

\noindent
{\bf Definition 9: (Complexity)}
{\em  The complexity of a material is defined as the dimension $M$ of its thermodynamic state space.}

\noindent
This quantity $M$ directly reflects the capability of a material to store information. For one-component gas systems, $M=1$, indicating that the system possesses essentially a single thermodynamic degree of freedom. Such a system exhibits no hysteresis and is therefore unusable for memory medium. By contrast, for solids, $M$ is generally the order ${\cal O}(N_{\rm at})$, indicating that solids possess a vastly greater capacity for storing information. A silicon crystal is complex enough even though the crystallographic structure is simple enough.
\footnote{The exact dependence of $N_{\rm at}$ on the number of configurations $N_{\rm c}$ is uncertain. Using some type of simulations, Kauffman showed that it scales as $\sqrt{N_{\rm at}}$  \cite{Kauffman97}. The present author is unsure whether this square-root dependence has universality beyond the validity of model. For the present purpose, however, it suffices to understand that $N_{\rm c}$ is as large as ${\cal O}((N_{\rm at})^{n})$ but far smaller than the number of microscopic states of the order ${\cal O}(e^{N_{\rm at}} )$.} 
In principle, any solid can serve as a memory material. 
The basic operation of modern electronic logical devices is based on the binary logic, which relies on bistable electronic states \cite{Landauer61}. Such bistability can be modeled by two equilibrium positions of a defect atom. A well-known example is the so-called $DX$ center in the compound semiconductor GaAlAs \cite{Mooney90}. If a crystal possesses $N_{d}$ such defect centers, it possesses $2^{N_{d}}$ possible equilibrium states and can therefore store an enormous amount of data. 

Defect-related metastable states are not unique to semiconductors; they are present in virtually all solid. Therefore, it is not surprising that memory effects are observed even in amorphous materials \cite{Vincent07,Kovacs79}. Of course, practical applications require much more than the mere existence of multiple stable states. Factors such as controllability, reproducibility, and switching speed impose severe restrictions, which is why only a limited number of materials are suitable for technological memory devices.
Biological systems provide particularly sophisticated mechanisms for exploiting the detailed arrangements of molecules. A large number of degrees of freedom, represented by $M$, can be manipulated through biochemical reactions. Among these, catalytic reactions by enzymes play a central role because they operate close to reversible conditions. The significance of such reversible transitions in information processing is discussed in Sec.~\ref{sec:Inf-Proc}.

\section{When does configuration entropy contribute to thermodynamic entropy?}
\label{sec:ConfS}
In Sec.~\ref{sec:FC}, we showed that the configuration entropy of solids is a frozen state variable and therefore does not present a current property of the material. Nevertheless, the thermodynamics literature contains many examples in which the mixing entropy of different configurations is treated as a genuine contribution to the entropy, seemingly supporting the validity of Gibbs entropy $S_{\rm G}$. To clarify this apparent inconsistency, it is necessary to distinguish carefully between mixing and transition, which is explained below.

\subsection{Mixing entropy of gases}
\label{sec:mix-gas}

\begin{figure}[htbp]
    \centering
    \includegraphics[width=160 mm, bb=0 0 800 160]{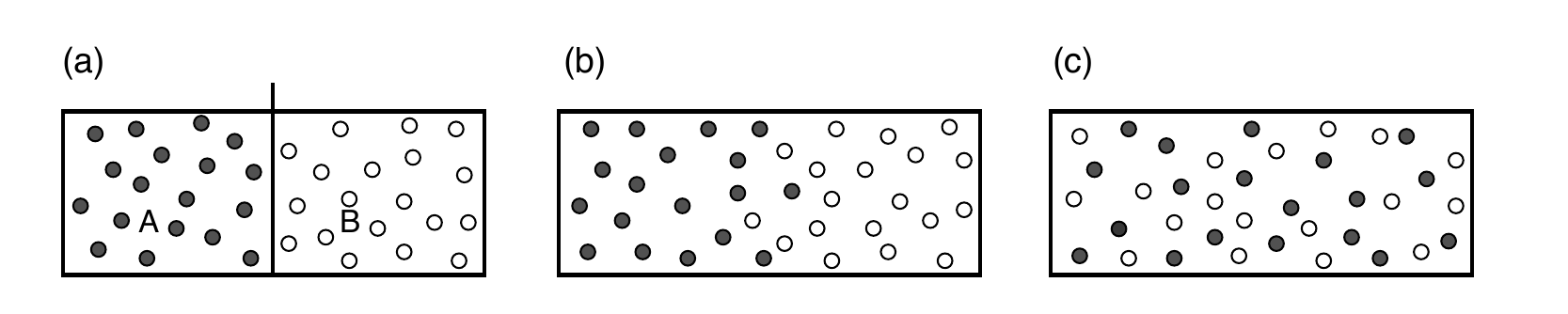} 
  \caption{Mixing of two gases, $A$ and $B$, in a container. (a) Initially, the two gases are separated by an internal wall. (b) Immediately after the wall is removed, the gases remain largely separately. (c) After sufficient time has elapsed, a homogeneous mixture is established.
  } \label{fig:mix-gas}
\end{figure}

Let us first examine the mixing entropy of multicomponent gases in the present framework, because it is well explained in standard textbooks. The mixing entropy $S_{\rm mix}$ of a multicomponent gas is given by
\begin{equation}
S_{\rm mix} = - k_{\rm B} N \sum_{i} x_{i} \ln x_{i},
\label{eq:mixS}
\end{equation}
where $x_{i}$ is the mole fraction of the $i$-th component gas. 
The appearance of Eq.~(\ref{eq:mixS}) is exactly the same as Gibbs entropy, Eq.~(\ref{eq:Gibbs-S}). 
Nevertheless, the mixed state should not be interpreted as an ensemble of different configurations.
A mixing process of two gases, $A$ and $B$, is illustrated in Fig.~\ref{fig:mix-gas}. The entropy of the entire system increases by $S_{\rm mix}$ by removing the internal wall. 
Immediately after removing the wall, the system is in a nonequilibrium state, corresponding to state (b) in the figure. After sufficient time has elapsed, the distributions of both gas species become uniform, and thus the mixed gas reaches the equilibrium state, corresponding to state (c) in the figure. 
The mixed state (c) is one and only one equilibrium state in this process, as Postulate 1 claims. Therefore, for this state, $S_{\rm B} = S_{\rm G}$. An unmixed state (b) is a nonequilibrium state and should not be included in the ensemble average in Eq.~(\ref{eq:Gibbs-S}). The separated state (a) is an equilibrium state only in the presence of the constraint. Although atom distributions of (a) and (b) appear almost the same, their thermodynamic meanings are different. In Eq.~(\ref{eq:mixS}), the summation should not be considered as the ensemble average of different configurations. The individual gas components are not separate equilibrium states. Rather, Eq.~(\ref{eq:mixS}) describes the entropy of a {\it single} equilibrium state containing multiple chemical species.

Abruptly removing a constraint causes a relaxation process, which is essentially an irreversible process. This results in a discontinuous change in the state variable $\Delta X$. In such a case, calorimetric measurement based on Eq.~(\ref{eq:defS}) does not give the correct entropy change $\Delta S$, since $Q=0$. To obtain the correct value $\Delta S$, this mixing process should be replaced by an equivalent reversible one, in which the state variable $X$ changes continuously. A standard example is the use of semipermeable membranes, whereby the partial volume change $\Delta V_{i}$ of each component is controlled through external work $W$, meaning that $V_{i}$ varies continuously. This technique is well established in the standard course of thermodynamics \cite{Cengel}.

\subsection{Configuration entropy in alloys and glasses}
\label{sec:ConfAlloy}
Now, we turn to the case of solids. In the thermodynamics literature, the concept of mixing entropy $S_{\rm mix}$ is invariably used for solids through an ensemble of different configurations. 
Typical examples include random alloys, defective crystals, and glasses: for the glass transition, extensive studies on the configuration entropy have bee reported \cite{Goldstein73,Berthier19a}. The term ``high-entropy alloy (HEA)" itself derives from this interpretation of configurational mixing. 
It should be noted, however, that configuration entropies in solids are inferred from theoretical investigations. As far as the author knows, there is no direct experiment showing such high entropy for HEA as expected from the configuration entropy \cite{Vaodya19,HEA-Zhang20,Haas18}. In experiment, the entropy of an HEA is determined from measurements on a single sample; no experiment performs the ensemble average over all samples. A single sample occupies only one configuration and accordingly the configuration entropy associated with that sample is expected to be $S_{c}=0$. 
This observation raises an important question: if a nonzero configurational entropy is deduced experimentally, what is the physical origin of that quantity? Resolving this issue requires a careful analysis of the experimental procedures by which the entropy is determined.

As discussed for the gas case above, mixing processes are irreversible processes and therefore the mixing entropy $S_{\rm mix}$ cannot be measured directly by calorimetric methods. Furthermore, special techniques such as insertion of semipermeable membranes cannot be applied to solids either. 
The concept of mixing entropy $S_{\rm mix}$ in solids is, however, validated through equilibrium condition. Consider a solute $A$ dissolved in a liquid solution $B$. The dissolving reaction is expressed as,
\begin{equation}
A^{\rm (atom)} + B^{\rm (liq)} \rightarrow A:B^{\rm (liq)}.
\label{eq:solute}
\end{equation}
At equilibrium, the concentration of dissolved atoms, $n_{A}^{\rm eq}$, can be obtained as
\begin{equation}
n_{A}^{\rm eq} = n_{B} \exp ( -\Delta H_{d}/k_{\rm B}T),
\label{eq:EqConc}
\end{equation}
where $\Delta H_{d}$ is the dissolution enthalpy of $A$ and $n_{B}$ is the density of the solvent $B$. 
Equation (\ref{eq:EqConc}) is derived from the equilibrium condition that the Gibbs free energies before and after dissolution, $G_{i}$ and $G_{f}$, are equal. In the expression $G_{f} = H_{f} -T S_{f}$, the entropy term contains the contribution $k_{\rm B} \ln (n_{B}/n_{A})$, which corresponds to the mixing entropy. This expression is exact for regular liquids and remains approximately valid for general liquids (\cite{Prigogine99}, Chap.~8).
Conventionally, the term $k_{\rm B} \ln (n_{B}/n_{A})$ may be interpreted as indicating a mixing of configurations, if the ratio $n_{B}/n_{A}$ is interpreted as the number of configurations $W_{\rm c}$: this way of interpretation is found in a textbook (\cite{Sachs06}, p.~62). In this picture, one might say that a solute atom $A$ experiences all possible ``configurations" in the liquid. 
\footnote{In fact, there are studies to decompose the entropy of a liquid in the form of Eq.~(\ref{eq:S=vib+conf}) in connection to the glass transition. See \cite{Alvarez-Donado20,Richert98,Johari00,Sastry01,Smith17,Han20}.}
However, this way of interpretation is inconsistent with the precise definition of configuration given in Definition 5. Instantaneous atom positions, ${\bf R}_{j}(t)$, in a liquid cannot be a state variable, and therefore the set $\{ {\bf R}_{j}(t) \}$ does not constitute a configuration in the thermodynamic sense. The mixing entropy appearing in reaction~(\ref{eq:solute}) is the entropy of a single equilibrium state of the mixed liquid, as for a mixed gas. After all, we see that for liquids only the total entropy is the physically meaningful quantity and decompositions such as in Eq.~(\ref{eq:S=vib+conf}) are merely conventions in order to explain a particular aspect of the total entropy \cite{Shirai-EntropyLiquid25}. 

As an extension of the liquid case, the equilibrium concentration of impurities, $n_{A}^{\rm eq}$, can also be considered for solids as an approximate description. Indeed, $n_{A}^{\rm eq}$ often provides a useful estimate of the solubility limit of an impurity $A$ in a host crystal $B$.
In this context, the reaction (\ref{eq:solute}) is reread as follows: $A$ represents the impurity atom and $B$ presents the host crystal. 
A crucial difference from the liquid case is that the solid solution state $A:B^{\rm (sol)}$ possesses a definite configuration $K$. The impurity atom A occupies a particular lattice site, and each sample therefore realizes only one specific configuration. The realized configuration is selected by chance during solidification and, once formed, does not transform to other configurations.
If the dissolution process is to be treated as reversible, the system must be able to explore all relevant configurations $\{ K \}$. This requirement is naturally satisfied in liquids. In solids, however, transitions between different configurations are generally frozen out. Consequently, the validity of applying the configurational entropy concept to solids requires further examination.

The requirement of reversibility can be satisfied through equilibrium between the liquid and solid phases. Near the melting temperature $T_{m}$, the reaction (\ref{eq:solute}) proceeds in both the forward and backward directions with equal probabilities. As a result, impurity atoms repeatedly enter and leave the solid phase. Through repeated forward/backward reactions, the system is able to explore the possible configurations $\{ K \}$ of impurity atom $A$ in solid $B$:
\begin{equation}
K \leftrightarrow K' \leftrightarrow K'' \leftrightarrow \cdots,
\label{eq:eqamongK}
\end{equation}
Only when transitions among different configurations occur freely in this manner can reversibility be established. This is precisely the meaning of equilibrium at the transition. All accessible configurations become activated, and genuine mixing of configurations takes place. Consequently, the configuration entropy, $S_{\rm c} = k_{\rm B} \ln W_{\rm c}$, becomes a real state variable and can be calculated using Boltzmann entropy $S_{\rm B}$. 
When temperature is lowered below $T_{m}$, atom motion becomes strongly restricted and the crystal freezes in a particular configuration. The configuration entropy $S_{c}$ (and therefore mixing entropy $S_{\rm mix}$) no longer represents a current property of the solid and instead becomes a frozen state variable. In this regime, $S_{\rm mix}$ is a quantity evaluated on the state space that existed prior to freezing, in the same sense as residual entropy.

In reality, the equilibrium concentrations predicted by Eq.~(\ref{eq:EqConc}), are never observed exactly in solids. Transitions (\ref{eq:eqamongK}) require a finite time of completion, regardless of how short that time may be. The characteristic time of transition is given by the relaxation time, $\tau_{KK'}$, which is determined by the energy barrier separating configurations $K$ and $K'$, as in Eq.~(\ref{eq:tau}). In experiments, cooling is always carried out on a laboratory timescale. Consequently, the system can access only a limited subset of of the possible configurations $\{ K' \}$ before atom rearrangement becomes frozen. As a result, the impurity concentration $n_{A}$ generally depends on the cooling rate.
In addition, the finite diffusivity of impurity atom $A$ gives rise to spatially inhomogeneous distributions of $A$ throughout the host crystal $B$ during cooling. As the temperature decreases, atom diffusion becomes progressively slower. At sufficiently low temperatures, atom migration effectively ceases. Consequently, the obtained concentration $n_{A}$ becomes frozen and no longer follow the temperature dependence predicted by Eq.~(\ref{eq:EqConc}). This behavior stands in sharp contrast to the liquid equilibrium case.
Indeed, the success of modern semiconductor technology relies on the ability to control impurity concentrations precisely. Device manufacturers treat $n_{A}$ as a controlling parameter determined by processing conditions rather than as an equilibrium quantity uniquely specified by temperature.

All experimentally observed configuration entropies of solids, including the configuration entropy $S_{\rm c}$ of random alloys, are analogous to the case of impurity concentration discussed above. Their thermodynamic significance arises through the equilibrium relations such as Eq.~(\ref{eq:K-ChemEQ}), which are approximately valid when samples are synthesized near the melting temperature $T_{m}$. Once the alloy has solidified, however, the equilibrium constant $K_{\rm eq}$ becomes a frozen state variable and loses its status as a current thermodynamic property, since the chemical composition no longer varies with $T$. 
Many confusions about the configuration entropy of glasses can be resolved by keeping the present view: refer to a recent study \cite{Shirai22-res} for the residual-entropy issue and \cite{Shirai-SH-Liquids25} for the glass-transition issue.
There is, however, a case in which the configuration entropy is observed in experiment. The direct observation of residual entropy in a random alloy was reported by Eastman and Milner \cite{Eastman33}. They observed nonzero $S_{\rm c}$ far below $T_{m}$ by an electrochemical method, even though the measurement was performed only a single sample. To date, no generally accepted microscopic explanation has been established for how the configuration entropy was detected in that method. The present author conjectures that the essential feature of the observation may be ascribed to the liquid nature of the electronic subsystem. The electromotive force is measured through an electric current and the current occurs only because of its liquid behavior. This interpretation is given in Appendix B. Regardless of the interpretation of the electrochemical measurements, the calorimetric method gives merely a frozen value for the configuration entropy of solids.

\subsection{Information processing}
\label{sec:Inf-Proc}

The information of a sold is conveyed by the state variables $\{\bar{\bf R}_{j}^{K} \}$ of the solid. Although the thermodynamic state space has a large dimension $3N_{\rm at}$, a particular sample occupies only one configuration $K$. The other configurations $\{ K' \}$ ($K' \neq K$) are frozen out in that sample, which appears to severely limit the potential of a solid as an information source. 
In reality, however, a single silicon chip is capable of processing a large amount of information. The essential feature that makes a silicon chip electronically versatile is its ability to undergo transitions among many equilibrium states---an $M$-bits memory chip possesses $2^{M}$ equilibrium states. The change in equilibrium states is made possible by external work.
Each equilibrium state is protected against spontaneous collapse by an energy barrier. To overcome this barrier, an external excitation, such as applying electric field is used. This is a direct consequence of Definition 1: for any equilibrium state, it is impossible to change it into another equilibrium state without external work (\cite{Gyftopoulos}, p.~64).

Information processing in electronic devices can be viewed as switching between equilibrium states. The energy input is required to overcome the energy barrier separating these states. In the literature, the minimum energy required for computation is often discussed in the context of repeated memory usage: namely, the minimum energy dissipation associated with resetting memory states, as expressed by Laundauer's erasure principle \cite{Landauer61,Keyes70,Bennett82,Leff-Rex2}. 
The minimum energy of $k_{\rm B}T \ln 2$ per bit is deduced from an entropy increase generated in the erasing process. 
The present study, however, arrives at a similar conclusion by a more direct argument. Work is required to overcome the energy barrier, $E_{b}$. Thus, the energy consumption can be decreased by reducing $E_{b}$. However, there is an intrinsic limitation for reducing $E_{b}$. If $E_{b}$ is less than  $k_{\rm B}T$, the state containing information spontaneously collapses. This sets the minimum energy consumption for information processing as $k_{\rm B}T$ per degree of freedom. 
This interpretation follows directly from the requirement of thermodynamic stability and is therefore more suitable than arguments based on Landauer's erasure principle.

Similar considerations can also be applied to biological systems.
A fundamental question in biology is why diverse individuals (phenotypes) emerge from a single genotype originating from the same fertilized cell. If the central dogma alone completely determined the outcome, one might expect identical individuals---in the sense of the same chemical components---to be produced. In reality, we observe that a single genotype produces a variety of phenotypes. In the present context, these phenotypes may be viewed as corresponding to different equilibrium states. Strictly speaking, living systems are open systems, and the concept of equilibrium states is not directly applicable. However, in restrictive conditions, a steady state $A^{J}$ maintained by an energy flow $J$ can be approximately regarded as equilibrium state $A^{0}$. A detailed discussion of this correspondence will be given elsewhere. For the present purpose, however, the following qualitative argument is sufficient. 

A principle for determining steady states under an imposed energy flow is provided by Prigogine's minimum entropy-production principle \cite{Prigogine67,Prigogine99}: for a given energy flow $J$, the system adopts a steady state $A^{J}$ that minimizes the entropy production. From the limit $J \rightarrow 0$, it follows from continuity that $A^{J}$ approaches $A^{0}$, which corresponds to an equilibrium state. It is therefore reasonable to expect that a steady state $A^{J}$ is present in the vicinity of each equilibrium state $A^{0}$. This means that the number of possible steady states is comparable to the number of possible equilibrium configurations $K$.
From this perspective, it is not surprising that a variety of individuals can develop from a single genotype, since various external inputs $J$ are received at every stage of the emergence process, thereby guiding the system toward different steady states.

An important point is that many such changes of states are brought about by transitions between microscopic structures $\{\bar{\bf R}_{j}^{K} \}$. When an Escherichia coli (E.~coli) is put in a glucose-deficient environment, the E.~coli tries to produce an appropriate protein $P_{1}$ to catalyze breaking larger sugar molecules. The production of $P_{1}$ is regulated by another protein $P_{2}$. An efficient regulatory mechanism is provided by a special region of DNA, called the operator, adjacent to the  main region containing the structural information of $P_{1}$ (\cite{Life-9}, Sec.~10.4). 
Under normal conditions, $P_{2}$ binds to the operator, which suppresses transcription of the genetic information for $P_{1}$ into RNA. However, in a glucose-deficient environment, the three-dimensional structure of $P_{2}$ is deformed, causing it to detach from the operator and accordingly initiates the transcription of the code $P_{1}$.
The functions as inhibitor and activator are mediated by different sites of the same protein $P_{2}$, which is known as {\it allosteric} enzyme. 

In this manner, depending on the environment, a living system changes the configuration $K$---in this case the number of proteins $P_{1}$ and $P_{2}$---in order to establish a new steady state. This process may be regarded as the activation of a frozen state variable---the configurational degree of freedom.
It is a common recognition that transitions in such biological processes often exhibit sigmoidal behavior, reminiscent of the switching characteristics of binary operations of electronic devices (\cite{Monod}, Chap. 4). The sigmoidal response suggests the existence of a threshold for the transition. The deflection point of the sigmoid may be interpreted as corresponding to an energy barrier. From this viewpoint, it is reasonable to model biological reactions as networks of binary logical units \cite{Kauffman95,Kaneko09}. Naturally, the cost of such structural changes must be supplied by the energy flow. Therefore, the biological evolution---often viewed as {\em order from disorder}, giving an impression of violating the second law of thermodynamics \cite{Schrodinger44,Kauffman95,Murphy-Oneill97}---does not violate the second law.
The process of overcoming energy barriers is primarily mediated by enzymatic catalysis.
In this manner, a living system exploits its abundant possible configurations $\{ K \}$. Under the presence of energy flow, the configurational degrees of freedom become active state variable. 
However, this activation does not mean to create mixing states but means to undergo transitions between different equilibrium states. Therefore, it is inappropriate to use an ensemble average over these different configurations to evaluate the entropy of a biological system.
If these states were truly mixed states, they could not carry useful information.

\section{Summary}
\label{sec:Conclusion}

At the beginning of this paper, several conceptual difficulties concerning entropy were raised. These include inconsistencies between order and entropy, randomness and order, and complexity and information. This study addresses these difficulties from one of the fundamental principles of thermodynamics: entropy is a state function and must therefore be uniquely determined by a thermodynamic state (Sec.~\ref{sec:thermodynamic-entropy}). 
The analysis of Sec.~\ref{sec:Origin} shows that these difficulties are, at a deep level, related to the long-standing issues in thermodynamics, such as the entropy attribution problem, hysteresis, and the third-law issues. Accordingly, resolving these conceptual difficulties requires simultaneously solving these long-standing issues, which is the reason why these difficulties have not long been resolved. The essential problem with previous theories is their ambiguous treatment of the equilibrium state. In particular, the conventional two-variables assumption has long obscured the resolution. As shown in this study, the two-variables assumption is incompatible with the thermodynamic behavior of solids.

Accordingly, the present study begins by defining thermodynamic equilibrium without invoking the concept of state variables, thereby avoiding the circular argument. This was done by Gyftopoulos and Beretta (Sec.~\ref{sec:EQ}). On this basis, the unambiguous definition of state variable has been achieved (Definition 3). The essential characteristic of state variable is its invariance under time averaging, irrespective of whether the quantity is microscopic or macroscopic (Definition 3). From this, the state variables of a solid are deduced as the time-averaged atom positions $\{ \bar{\bf R}_{j} \}$ (Corollary 3). Thus, a variety of static states of a solid exist for a given $T$ and $V$, including defect states. These states are genuine equilibrium states and are therefore completely characterized by their state variables, irrespective of the past history by which they were reached (Corollary 1). Consequently, every equilibrium state possesses a unique entropy determined by the FRE, Eq.~(\ref{eq:S-FREsolid}). 

The essence of state variables is their time-averaging invariance (Definition 3). This property also underlies the notions of order in materials (Corollary 8) and information of materials (Corollary 9). Entropy represents the degree of thermal fluctuations associated with the state variables $\{ X_{j} \}$. Hence, entropy is distinct from both the order and the information of materials. Likewise, entropy should not be regarded as a measure of material complexity. Rather, it is suitable to ascribe the complexity to the dimensionality $M$ of the thermodynamic space of a material (Definition 9).

\begin{figure}[htbp]
    \centering
     \includegraphics[width=120 mm, bb=0 0 600 450]{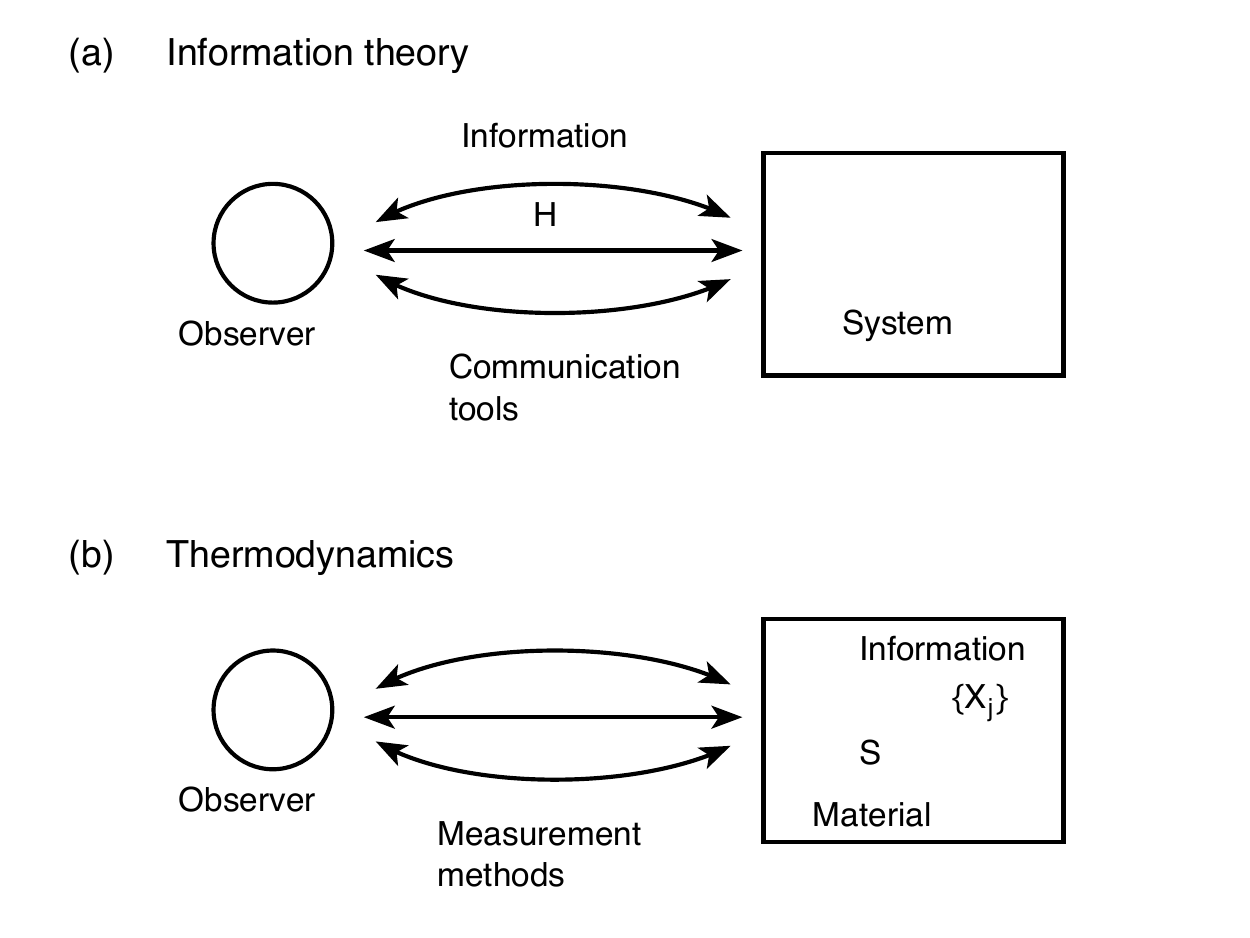} 
  \caption{(a) Information in information theory. The information data are carried through a channel between a system and an observer. (b) Information in thermodynamics. The information is a property of a material described by state variables $\{ X_{j} \}$.
  } \label{fig:Info-scheme}
\end{figure}

Figure~\ref{fig:Info-scheme} summarizes the fundamentally different roles of information in information theory and thermodynamics. In information theory, typically telecommunication theory, the information data are carried through the communication channel between a system and an observer. The information available to the observer depends on the observation and communication processes. The information-theoretic entropy $H$ quantifies this information. Accordingly, the entropy assigned to a system may change as new information becomes available through observation. This may be similar to that work and heat depend on processes but are not properties of a system.
In thermodynamics, on the other hand, the information is expressed by the state variables $\{ X_{j} \}$ of a material. Therefore, the information of a material is independent of the measurement methods by which it is obtained. Because the entropy $S$ of a material is also expressed by these state variables, it is independent of the measurement.

Although the requirement of the time-average invariance is satisfied, there is a special class of state variables, that is, frozen state variable, which becomes constant because the associated constraint is so strong that it cannot vary with variations of other variables (Sec.~\ref{sec:FC}). Hence, frozen state variables do not appear in the FRE explicitly. All the remaining state variables are active state variables.
By definition, a state of a given solid is specified by the active state variables, which are specified by the current configuration $K$ alone. All other configurations are frozen state variables, and should not be included in the current value of entropy. Therefore, the entropy of any material must converge to zero when $T$ approaches 0 (Postulate 2).

The configuration entropy of solids is associated with frozen variables, which are remnants of the configurational degrees of freedom that were active in the high-temperature liquid state  (Sec.~\ref{sec:ConfS}). Accordingly, configuration entropy of a solid is not the property of the present equilibrium state.
However, transitions between different configurations activate previously frozen state variables. Such activation occurs during phase transitions or whenever external work supplies sufficient energy to overcome the associated constraints.

Information processing in electronic devices is possible because a single solid possesses an enormous number of equilibrium states---not because many different samples of the same material exist. Transitions between these equilibrium states may be regarded as the activation of frozen configurations, driven by external supplied work. 
The same framework may provide a thermodynamic perspective on phenotypic diversity in living systems, in which metabolic energy activates transitions among equilibrium states.

\section*{Acknowledgement}
The author thanks Prof.~J.~Gielis (U.~Antwerp) for inviting me to workshop of Square Bamboos and the Geometree, which gave me an opportunity to extend my research in thermodynamics to biological science.

\section*{Appendix A: Cyclic processes and the emergence of rules}
Discussions of the origin of life inevitably encounter a circular argument resembling the classic egg-or-chicken problem: which appeared first, DNA or metabolism? \cite{Thaxton20,Kauffman95,Davies19}. The blueprint of living organisms is encoded in DNA, yet the origin of this information remains one of the most challenging questions in the life sciences.
In Sec.~\ref{sec:Information}, it was suggested that the particular correspondence between codons and amino acids may have been established by chance. Many researchers have argued that the probability of obtaining a biologically meaningful sequence through purely random events is vanishingly small \cite{Thaxton20,Kauffman95,Davies19}. However, this conclusion is not necessarily unavoidable.

Even if an initial chemical transformation, $A_{1} \rightarrow A_{2}$, occurs by chance, the situation can change qualitatively when a sequence of subsequent reactions, $\{ A_{i} \rightarrow A_{i+1} \}$  closes to form a cycle under an energy flow. Such a cycle occupies a special position in thermodynamics because it permits repeated and sustained operations. Cyclic processes provide a natural mechanics for maintaining stable steady states. If the cycle is reversible, it is the most energy efficient cycle like Carnot cycle. This feature is actually observed in biological metabolism, A well-known example is the Krebs cycle. Interestingly, the reverse Krebs cycle is also known and is believed to have played an important role at the early stage of biological evolution \cite{Lane22}.

The analogy between metabolism and heat engines deserves emphasis. A cycle of a heat engine is composed of expansion and contraction, despite that only the expansion step is useful for obtaining work. The reason why cyclic operation is indispensable for a heat engine is straightforward: an ever-expanding engine cannot be installed in an automobile (\cite{Cengel}, p.~251). A practical engine must repeatedly return to the initial state. The restoration of the initial state, which makes the process cyclic, is therefore essential. 
This seemingly trivial observation has consequences that extend far beyond engineering applications. Repeated operation of a cycle naturally produces stable steady states, and stable steady states provide a basis for the emergence of persistent rules. Lane has argued that metabolic cycles may have served as prototypes of the template mechanisms later embodied in DNA \cite{Lane22}. See also \cite{Duve97}. From a thermodynamic viewpoint, this proposal appears plausible. A particular correspondence between a cause and its effect may initially arise by chance, but repeated cycling can stabilize and preserve that correspondence, thereby transforming a chance event into an enduring rule.

\section*{Appendix B: Electromotive force in a random alloy}
Liquids do not possess microscopic stationary states (eigenstate in the quantum-mechanical words). This fact has significant consequence on the thermodynamic properties of liquids. A detailed discussion of this issue is given in a recent paper \cite{Shirai-SH-Liquids25}. As argued below, this liquid nature is eventually related to the electronic response of solids.

The standard theoretical description of electronic states in solids is provided by band theory, which treats electrons within an effective one-electron picture in a static potential field \cite{Ashcroft-Mermin}. Here, ``static'' means that the potential is given by the time-averaged atomic positions ${\bar{\bf R}_{j}^{K}}$ rather than the instantaneous positions ${{\bf R}_{j}(t)}$.
Within band theory, a set of single-particle eigenstates $\{ \varepsilon_{j}^{K} \}$ is obtained from the effective potential $V(\{ \bar{\bf R}_{j}^{K} \})$. 
These eigenstates determine the thermodynamic properties of the solid, including electrical conductivity and electromotive force.
By definition, eigenstates have infinite lifetimes, implying that an electron in a solid experiences the crystal potential at all positions $\{ \bar{\bf R}_{j}^{K} \}$.
This means that frozen configurations $K'$ ($K' \neq K$) cannot appear in the electronic properties of the individual sample, because one sample occupies only one active configuration $K$.

At this point, it should be recalled that the independent-particle description is only an approximation. In reality, the electrons in a solid constitute an interacting many-body system, which is often described as a Fermi liquid \cite{Nozieres64,K-Yamada93}. The one-particle eigenstates of band theory are valid only in an approximate sense. Each quasiparticle state is characterized by a finite lifetime $\tau_{j}$. The eigenstate $\varepsilon_{j}^{K}$ therefore has physical significance only within the timescale $\tau_{j}$.  During this period, an electron can travel only in a limited region of the solid.
Consequently, electrons located in different places of the sample experience different local portions of the total potential $V(\bf r)$. In a random alloy, the local atomic arrangement varies from place to place within a single sample. These local variations statistically represent the distribution of atomic arrangements that would be observed among different samples. As a result, different electrons effectively experience different local environments—or, in an approximate sense, different ``configurations''—within the same sample.


\begin{thebibliography}{159}
\bibitem [{not()}]{note-Sommerfeld}%
  \BibitemOpen
  \href@noop {} {}\bibinfo {note} {Sommerfeld said ``Thermodynamics is a funny
  subject. The first time you go through it, you don't understand it at all.
  The second time you go through it, you think you understand it, except for
  one or two points. The third time you go through it, you know you don't
  understand it, but by that time you are so used to the subject, it doesn't
  bother you anymore" \cite{Cengel}.}\BibitemShut {Stop}%
\end{thebibliography}

%

\end{document}